\documentclass[11pt]{article}
\usepackage{tikz} 
\usetikzlibrary{positioning}
\usepackage{geometry}
\usepackage{booktabs}
\usepackage{lineno}
\usepackage{multirow}
\usepackage{enumitem}
\usepackage{amsmath}
\usepackage{float}
\usepackage{graphicx} 
\usepackage{amssymb,amscd,amsmath,amsthm,color}
\usepackage[T1]{fontenc} 
\usepackage{subcaption}
\usepackage{enumitem}
\usepackage{mathtools}
\usepackage{tcolorbox}
\usepackage{verbatim}
\usepackage{mathrsfs}
\usepackage{physics}
\usepackage{mathtools}
\usepackage{slashed}
\usepackage{longtable}
\usepackage{jheppub}
\usepackage{hyperref}
\usepackage{cleveref}

\def\be{\begin{equation}}
	\def\bea{\begin{eqnarray}}
		\def\ee{\end{equation}}
	\def\eea{\end{eqnarray}}

\def\no{\nonumber}
\def\half{\frac{1}{2}}

\numberwithin{equation}{section}

\newcommand{\rmd}{d}

\newcommand{\diff}{d}

\newcommand{\jr}{J_{3R}}
\newcommand{\jl}{J_{3L}}

\usepackage{enumerate}
\usepackage{slashed, enumitem}
\usepackage{array}
\newcolumntype{P}[1]{>{\centering\arraybackslash}p{#1}}
\usepackage{anyfontsize}

		\author[a]{Kanhu Kishore Nanda,}
		\author[b]{P. Shanmugapriya,}
		\author[a]{Amitabh Virmani}
		\affiliation[a]{\it Chennai Mathematical Institute, H1 SIPCOT IT Park, Siruseri, Chennai, 603013, India\\  }
		\affiliation[b]{\it International Centre for Theoretical Sciences, TIFR Bengaluru, 560089, India \\}
		\emailAdd{kanhukishore@cmi.ac.in}
		\emailAdd{shanmugapriya.prakasam@icts.res.in}
		\emailAdd{avirmani@cmi.ac.in}

\abstract{Boruch, Emparan, Iliesiu, and Murthy recently discussed index saddles for 5D black strings, showing that the black string saddle admits a decoupling limit to a complex, finite-temperature BTZ~$\times~S^2$ saddle that computes the index of the dual CFT. In this paper, we pursue an analogous construction for the D1-D5-P black string. We construct a four-charge index saddle in the four-dimensional STU model as the BPS limit of the non-extremal four-charge black hole, and show that it exhibits the new form of attraction. We then uplift it successively to five and six dimensions, via the 4D-5D connection and a chain of string dualities, to obtain the gravitational index saddle for the D1-D5-P black string. We take a systematic decoupling limit of this index saddle and obtain the BTZ~$\times~S^3$ saddle that computes the index of the D1-D5 CFT.}
	
\title{Index saddle for the D1-D5-P black string and its decoupling limit}
	
\begin{document}
\maketitle
\section{Introduction}

The statistical explanation of black hole entropy is one of the central achievements of
string theory. For a large class of supersymmetric black holes, the microscopic count of
states at weak coupling, organized as a supersymmetric index, reproduces the
Bekenstein-Hawking entropy at large charges~\cite{Strominger:1996sh,Sen:2007qy}.
The supersymmetric index 
\begin{equation}
	\mathcal{I} \;=\; \mathrm{Tr} \, (-1)^F \, e^{-\beta H}
\end{equation}
is well-defined at any finite inverse temperature $\beta > 0$, because the supersymmetry
algebra pairs non-BPS states so that their contributions cancel, rendering the index
$\beta$-independent. At strong coupling, the same charges are carried by a supersymmetric
black hole solution of supergravity, and one expects the index to be reproduced by a
gravitational computation. However, this leads immediately to a basic tension: the
regular Lorentzian solutions for supersymmetric black holes are necessarily extremal,
existing only at zero temperature $\beta \to \infty$, while the index is defined at
finite $\beta$. Moreover, the Bekenstein-Hawking entropy counts bosonic and fermionic
states with the same sign, whereas the index counts them with opposite signs. The
precise gravitational mechanism that reconciles these two facts within a single framework
has long remained unclear.

A resolution was recently found through the identification of gravitational saddles
that contribute to the supersymmetric index. Finite-temperature saddle-point contributions were 
constructed for black holes in AdS \cite{Cabo-Bizet:2018ehj} and in flat space \cite{Iliesiu:2021are}. These saddles are smooth, complex 
Euclidean spinning black holes that are supersymmetric but not extremal. They are formally  finite temperature solutions saturating the BPS bound, and they impose periodic
boundary conditions on fermionic fields around the thermal circle, correctly implementing
the $(-1)^F$ insertion. 

A key feature of these solutions is the new attractor
mechanism \cite{Boruch:2023gfn}. At the north and south poles of the rotating Euclidean horizon, the scalar
moduli are attracted to values that depend only on the  charges of the
solution, not on the temperature or the asymptotic values of the moduli.  As a consequence, the on-shell action, and hence the
gravitational index, is independent of $\beta$, in precise agreement with the
$\beta$-independence of the microscopic index. These developments provide
the long sought after gravitational explanation of why the logarithm of the index equals the entropy for
supersymmetric black holes.

The program of finding gravitational index saddles was subsequently extended  by Boruch, Emparan, Iliesiu, and Murthy~\cite{Boruch:2025qdq} among other authors \cite{Anupam:2023yns, Hegde:2023jmp,  Adhikari:2024zif, Cassani:2024kjn, Hegde:2024bmb,  Adhikari:2025eok, Cassani:2025iix, Bandyopadhyay:2025jbc, Boruch:2025sie, Boruch:2025biv, Dharanipragada:2026dji};  see \cite{Cassani:2025sim} for a review. Boruch et al work  in
$\mathcal{N}=2$ supergravity obtained via a Calabi-Yau compactification of Type IIA-theory. They 
construct the five-dimensional index saddles by uplifting the four-dimensional index saddles via the M-theory circle. The choice of D6-brane charge $\Gamma^0$ of the
four-dimensional seed solution determines the topology of the uplift \cite{Gaiotto:2005gf}. $\Gamma^0 =
1$ yields a five-dimensional black hole saddle (upon taking an appropriate zoom-in limit) in asymptotically flat $\mathbb{R}^4
\times S^1_\beta$, while $\Gamma^0 = 0$ yields a five-dimensional black string saddle
wrapping the M-theory circle $S^1_M$, with asymptotic geometry $\mathbb{R}^3 \times S^1_M \times
S^1_\beta$.  An important observation made by Boruch et al for the black string was that as the temperature is lowered sufficiently  in the
flat space region, the solution admits a novel decoupling limit in which the AdS$_3$ throat takes the
form of a finite-temperature BTZ black hole. The finite-temperature BTZ black hole computes the index in AdS$_3$/CFT$_2$; it contributes to the index of the dual Maldacena-Strominger-Witten (MSW) CFT \cite{Maldacena:1997de}.
 
It is natural to pursue an analogous construction for the D1-D5-P black string. This string famously admits $AdS_3 \times S^3$ near-horizon geometry, and its microscopic physics is governed by
the two-dimensional D1-D5 CFT. The natural question is then, what is the gravitational saddle
in six dimensions that computes the index of the D1-D5 CFT?

In this paper, we answer this question. Our strategy is summarized schematically in Figure~\ref{fig:roadmap}. Boruch et al reached their five-dimensional black string by uplifting a four-dimensional seed solution along the M-theory circle. This is the brane system of three stacks of intersecting M5-branes carrying momentum, the system underlying the MSW SCFT. The brane content relevant for the D1-D5-P system is different: three stacks of M2-branes, the system that gives the Breckenridge-Myers-Peet-Vafa (BMPV) black hole in five dimensions. Reaching it requires instead the choice $\Gamma^0 = 1$, for which the same M-theory circle uplift yields a five-dimensional saddle in asymptotically flat space rather than a black string.

We therefore start in four dimensions and construct a new index saddle for a black hole carrying four independent charges in the STU model, obtained as the supersymmetric, finite-temperature limit of the known non-extremal, rotating solution. We verify that this saddle satisfies the new attractor mechanism. Uplifting this four-dimensional saddle along the M-theory circle with $\Gamma^0 = 1$, we recover the known five-dimensional Anupam-Chowdhury-Sen (ACS) index saddle~\cite{Anupam:2023yns} for the BMPV black hole. We uplift this index saddle to IIB 
supergravity.  The resulting
solution is a smooth, complex Euclidean black string in six dimensions (IIB theory on $T^4$), with periodic fermion boundary conditions around the thermal
circle. Taking the near-horizon limit of the
six-dimensional solution, we obtain the BTZ~$\times~S^3$ saddle.

\definecolor{roadblue}{rgb}{0.122,0.306,0.588}
\definecolor{roadbluefill}{rgb}{0.918,0.945,0.984}
\definecolor{roadgreen}{rgb}{0.180,0.490,0.196}
\definecolor{roadgreenfill}{rgb}{0.918,0.965,0.918}
\definecolor{roadgold}{rgb}{0.722,0.525,0.043}
\definecolor{roadgoldfill}{rgb}{1.0,0.973,0.882}
\definecolor{roadpurple}{rgb}{0.416,0.106,0.604}
\definecolor{roadpurplefill}{rgb}{0.953,0.910,0.984}
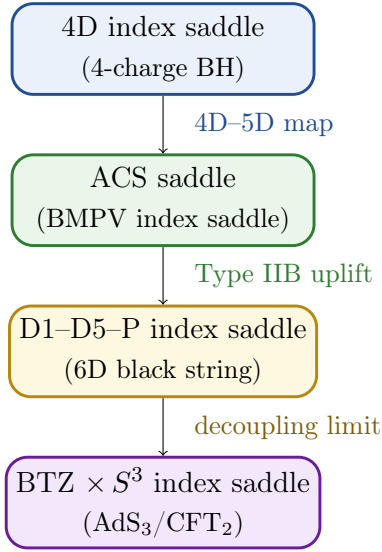
\begin{figure}[t]
	\centering
	\begin{tikzpicture}[
		box/.style={
			rectangle, rounded corners=8pt, draw, line width=1.1pt,
			minimum width=4cm, minimum height=1.2cm, align=center
		}
		]
		
		\node[box, fill=roadbluefill, draw=roadblue] (saddle4d) at (0,6)
		{ 4D index saddle\\ \small (4-charge BH)};
		
		\node[box, fill=roadgreenfill, draw=roadgreen] (saddleACS) at (0,4)
		{ ACS saddle\\ \small (BMPV index saddle)};
		
		\node[box, fill=roadgoldfill, draw=roadgold] (saddle6d) at (0,2)
		{ D1--D5--P index saddle\\ \small (6D black string)};
		
		\node[box, fill=roadpurplefill, draw=roadpurple] (saddleAdS) at (0,0)
		{ BTZ $\times\, S^3$ index saddle\\ \small (AdS$_3$/CFT$_2$)};
		
		\draw[->] (saddle4d) -- (saddleACS)
		node[midway, anchor=west, xshift=8pt, color=roadblue] {\small 4D--5D map};
		\draw[->] (saddleACS) -- (saddle6d)
		node[midway, anchor=west, xshift=8pt, color=roadgreen] {\small Type IIB uplift};
		\draw[->] (saddle6d) -- (saddleAdS)
		node[midway, anchor=west, xshift=8pt, color=roadgold!70!black] {\small decoupling limit};
		
	\end{tikzpicture}
	\caption{Schematic relation between the various index saddles discussed in this work.
		The four-dimensional index saddle uplifts via the 4D--5D map to the ACS
		saddle~\cite{Anupam:2023yns}, which in turn uplifts to a six-dimensional D1--D5--P black string
		saddle in Type IIB supergravity. The decoupling limit of
		this geometry yields a BTZ$~\times~S^3$ saddle that computes the index of the D1--D5 CFT.
	}
	\label{fig:roadmap}
\end{figure}

We note that, while this work was in progress, Larsen and Sharma~\cite{Larsen:2026sav} independently studied closely related solutions with asymptotic $AdS_3 \times S^3 \times T^4$ boundary conditions, reached either (i) by directly choosing four-dimensional multi-center data that produces them, or (ii) by taking a near-horizon limit of a non-extremal six-dimensional black string and then imposing the BPS condition. Our derivation instead proceeds via an explicit large-radius decoupling limit of the asymptotically flat D1-D5-P index saddle, extending to six dimensions the analogous treatment of~\cite{Boruch:2025qdq} in five dimensions.

The rest of the paper is organized as follows. In Section~\ref{sec:4charge}, we construct the four-dimensional index saddle for a four-charge black hole in the STU model: we review the relevant non-extremal, rotating solution (\S\ref{CC_non-BPS}) and take its supersymmetric limit (\S\ref{CC_bps}), recast the resulting BPS solution in the Bates-Denef two-center form (\S\ref{sec:batesdenef}), and show that it furnishes a further example of the new attractor mechanism (\S\ref{sec:newattractor}). In Section~\ref{sec:D1D5saddle}, we uplift this saddle to obtain the index saddle for the D1-D5-P black string: we review the Type IIB uplift of five-dimensional supersymmetric solutions and recast the ACS saddle for the BMPV black hole in Bena-Warner harmonic-function form (\S\S\ref{sec:typeIIBuplift}--\ref{sec:harmonicfunctions}), match these harmonic functions to the attractor charges of the four-dimensional saddle via the 4D-5D connection (\S\ref{sec:4d5dconnection}), and assemble the resulting six-dimensional index saddle (\S\ref{sec:indexsaddleD1D5}). In Section~\ref{sec:decoupling}, we take a systematic large-radius decoupling limit of this asymptotically flat solution and obtain the BTZ$~\times~S^3$ saddle that computes the index of the D1-D5 CFT. We close with a discussion of open questions in Section~\ref{sec:conclusions}. Two appendices present related calculations of interest. Appendix \ref{5d6dbpsapp} constructs the six-dimensional D1-D5-P index saddle by applying the supersymmetric limit directly on the  non-extremal D1-D5-P black string. Appendix \ref{4d5dapp} provides another related check by extending the Emparan-Maccarrone \cite{Emparan:2007en} zooming-in procedure to the four-charge four-dimensional black hole of Section~\ref{sec:4charge}, recovering the non-extremal five-dimensional black hole.

\bigskip 
\noindent
\textbf{Note added:} After this paper appeared, Georgescu, Murthy, and Svesko~\cite{Georgescu:2026uhv} independently obtained overlapping results on the D1-D5-P index saddle and its decoupling limit.

\section{The index saddle for a four-charge black hole}
\label{sec:4charge}

In this section we construct the gravitational saddle that computes the supersymmetric index of a black hole carrying four independent charges. We work in the STU model, that is, $\mathcal{N} = 2$ supergravity coupled to three vector multiplets. This four-charge black hole is a natural next case to study beyond the cases already known in the literature: it has enough charges to be generic, while still being simple enough that its non-extremal, rotating solution is well studied in the literature.

We build the saddle in four steps. In Section \ref{CC_non-BPS} we start from the general non-extremal, rotating, four-charge solution of the STU model. In Section \ref{CC_bps} we take its supersymmetric limit: the mass is fixed in terms of the charges, but the solution remains at finite temperature rather than degenerating to the extremal horizon. In Section \ref{sec:batesdenef} we rewrite this BPS, non-extremal solution in the Bates-Denef two-center form. This rewriting is what exposes the structure we are after: in Section \ref{sec:newattractor} we show that the charges at the two centers are exactly the ones fixed by the new attractor mechanism for the index. This final solution is already discussed in the literature \cite{Boruch:2025qdq}; our construction of taking the BPS limit of the non-extremal solution is novel.

\subsection{Non-extremal four charge black hole} \label{CC_non-BPS}

The STU model Lagrangian is most efficiently fixed using the $\mathcal{N} = 2$ prepotential formalism. We take the prepotential to be 
\be
F(X) = - \frac{X^1 X^2 X^3}{X^0},
\ee
a homogeneous function of degree two in the four complex scalars $X^M$, $M = 0,1,2,3$. These are projective coordinates: only three of the four complex degrees of freedom are physical, the remaining one being removed by a choice of gauge. Defining $F_M \equiv \partial F/\partial X^M$, the generalized K\"ahler potential is
\be
e^{-\mathcal{K}(X, \bar{X})} = i \, (\bar{X}^M F_M - \bar{F}_M X^M),
\ee
and we fix the residual scaling freedom by working in the D-gauge, 
\be
e^{-\mathcal{K}(X,\bar X)} = 1,
\ee 
together with $\bar{X}^0 = X^0$. This leaves three complex, i.e.\ six real, physical scalars. The two-derivative action is completely determined by the prepotential $F$, so that all physical quantities can be expressed in terms of $X^M$ and $\bar{X}^M$. In particular, the Lagrangian takes the form
\be \label{4D_PP_Lag}
\mathcal{L}_4 = R \star_4 1 - 2 G_{MN} \star_4 \rmd X^M \wedge \rmd \bar{X}^N + \half \operatorname{Im} \, \mathcal{N}_{MN} \, \check{F}^M \wedge \star_4 \check{F}^N + \half \operatorname{Re} \, \mathcal{N}_{MN} \, \check{F}^M \wedge \check{F}^N,
\ee
where $G_{MN}$ is the K\"ahler metric and $\mathcal{N}_{MN}$ is the period matrix governing the gauge kinetic terms,
\bea
G_{MN} &=& \frac{\partial^2 \mathcal{K}}{\partial X^M \, \partial \bar{X}^N}, \\
\mathcal{N}_{MN} &=& \bar{F}_{MN} + \frac{N_{MK} \, X^K \, N_{NL} \, X^L}{X^M \, N_{MN} \, X^N}, \qquad N_{MN} = F_{MN} - \bar{F}_{MN}.
\eea

To exhibit the bosonic content explicitly, we go to the duality frame singled out by the prepotential above. The bosonic field content is then a 4D metric $g_{\mu \nu}$, the four $U(1)$ gauge fields $A^0, A_i$, three dilatons $\varphi_i$ and three axions $\chi_i$. The gauge field strengths are given by $\check{F}^0 = dA^0$ and $\check{F}_i = dA_i$, and the Lagrangian reads
\begin{align} \label{tilde123Lagrangian}
	\mathcal{L}_4 & = R \star_4 1 - \half \sum_{i = 1}^3 (\star_4 \rmd \varphi_i \wedge \rmd \varphi_i + e^{2 \varphi_i} \star_4 \rmd \chi_i \wedge \rmd \chi_i) - \half e^{-\varphi_1 - \varphi_2 -\varphi_3} \star_4 \check{F}^0 \wedge \check{F}^0 \no \\
	& \quad - \half \sum_{i = 1}^3 e^{2 \varphi_i - \varphi_1 - \varphi_2 - \varphi_3} \star_4 (\check{F}_i - \chi_i \, \check{F}^0) \wedge (\check{F}_i - \chi_i \, \check{F}^0) - \chi_1 \chi_2 \chi_3 \, \check{F}^0 \wedge \check{F}^0 \no \\
	& \quad + (\chi_1 \chi_2 \, \check{F}_3 + \chi_2 \chi_3 \, \check{F}_1 + \chi_3 \chi_1 \, \check{F}_2) \wedge \check{F}^0 - \chi_1 \, \check{F}_2 \wedge \check{F}_3  - \chi_2 \, \check{F}_3 \wedge \check{F}_1 - \chi_3 \, \check{F}_1 \wedge \check{F}_2,
\end{align}
with the dictionary
\be
\frac{X^i}{X^0} = \chi^i + i \exp[-\varphi_i].
\ee

A general non-extremal black hole solution of this theory is a 10-parameter family, with mass and rotation parameters $(m, a)$ and four each of electric and magnetic charges. In this paper we restrict to the six-parameter solution obtained by setting the electric charge of $A^0$ and the magnetic charges of $A_1, A_2, A_3$ to zero. This leaves only the mass $m$, the rotation parameter $a$, the magnetic charge of $A^0$, and the electric charges of $A_1, A_2, A_3$. In the parametrisation of Chow and Comp\`ere \cite{Chow:2013tia, Chow:2014cca}, this is precisely the rotating, non-supersymmetric, non-extremal four-charge solution. The NUT parameter and all ``electric'' parameters $\delta_I$ are turned off, while all ``magnetic'' parameters $\gamma_I$ are turned on. Four-charge black holes in string theory have a long history, starting with the seminal work of Cveti\v{c} and Youm \cite{Cvetic:1996kv} and including the presentation of the  four-charge rotating solution of Chong, Cveti\v{c}, L\"u and Pope \cite{Chong:2004na}. 

Throughout this section we follow the notation of Chow and Comp\`ere \cite{Chow:2013tia, Chow:2014cca}, whose gauge field $A^4$ is our $A^0$ and $\widetilde A_i$ are our $A_i$. To keep the presentation as close as possible to theirs, and to avoid a cumbersome change of notation, we parameterise the magnetic charge of $A^0$ by $\gamma_4$ rather than relabeling it $\gamma_0$. This is admittedly a little awkward; but the solution and the fields built from it are already complicated enough that we find it best to match the literature directly rather than introduce a second parallel set of notation.

The black hole mass, angular momentum and the charges carried by the non-extremal black hole are given by\footnote{Throughout, $c_I \equiv \cosh\gamma_I$ and $s_I \equiv \sinh\gamma_I$, and for any subset of indices we write $c_{I_1 \cdots I_k} \equiv c_{I_1} \cdots c_{I_k}$ and $s_{I_1 \cdots I_k} \equiv s_{I_1} \cdots s_{I_k}$ for the corresponding products of these hyperbolic functions; e.g.\ $c_{1234} = c_1 c_2 c_3 c_4$ and $s_{13} = s_1 s_3$.},\footnote{The sub- and super-script $\mathrm{ne}$ on $p^0_\mathrm{ne}$ and  $q_i^\mathrm{ne}$ refers to charges for the \textit{non-extremal} black hole.}
\begin{align}
	& G_4 M = \frac{m}{4} \sum_{I=1}^{4} \cosh{2\gamma_I} = \frac{m}{4} \, \sum_{I=1}^{4} (2 c_I^2 - 1), \\
	& G_4 J = m a \, (c_{1234} - s_{1234}), \\
	&p^0_\mathrm{ne} = 2 m \sinh{\gamma_4} \cosh{\gamma_4} = 2 m s_4 c_4, \\
	&q_i^\mathrm{ne} = 2m \sinh{\gamma_i} \cosh{\gamma_i} = 2m s_i c_i.
\end{align}
The metric of the non-extremal solution takes the  form
\bea \label{CC_non-ext}
ds^2 &=& - \frac{r^2-2m r + a^2 \cos^2{\theta}}{W} (dt + \omega_3)^2 \no \\
&& + \ W \ \Bigg( \frac{dr^2}{r^2-2m r + a^2} + d\theta^2 + \frac{r^2-2m \, r + a^2}{r^2-2m r + a^2 \cos^2{\theta}} \sin^2{\theta} \, d\phi^2 \Bigg),
\eea
where 
\be 
W^2 = (r^2-2m \, r + a^2 \cos^2{\theta}) \Big[ r^2 + 2mr \Big(1 + \sum_I s_I^2 \Big) + a^2 \cos^2{\theta} + 4m^2 C \Big] + 4m^2 \hat{L}^2,
\ee
with 
\bea 
C & =& 1 + \sum_I s_I^2 + \sum_{I<J} s^2_I s^2_J - (c_{1234}-s_{1234})^2, \\
\hat{L} &=& (c_{1234}- s_{1234}) \, r + 2m s_{1234}.
\eea
The one-form $\omega_3$ is\footnote{We note that Eq.~(7.14) and Eq.~(7.16) of \cite{Chow:2014cca} are incorrect. In the context relevant to our work, the correct expressions are obtained by setting $\delta_I = 0$ in their Eq.~(5.17). In the context of \cite{Chow:2014cca} itself, however, the correct expressions for Eq.~(7.14) and Eq.~(7.16) are instead obtained by setting $\gamma_I = 0$ in Eq.~(5.17). Our Eq.~(2.14) matches with Eq.~(35) of \cite{Chong:2004na} when all our $\gamma_I$ are instead replaced with $\delta_I$.}
\be 
\omega_3 = \frac{2 m a \, \hat{L} \sin^2{\theta}}{r^2 - 2 m r + a^2 \cos^2{\theta}} d\phi.
\ee

The four-dimensional gauge fields are given as follows
\bea 
A^0 &=& - \zeta^4 (dt + \omega_3) - A^4_{3D}, \\
A_i  &=& \widetilde{\zeta}_i (dt + \omega_3) + (\widetilde{A}_i)_{3D}, \qquad i = 1,2,3
\eea
with the three-dimensional gauge fields given as
\bea 
(\widetilde{A}_i)_{3D} &=& -\frac{2 a m \, \sin^2{\theta}}{(r^2 - 2m r + a^2 \cos^2{\theta})} \, \Big[ \Big( \frac{s_i}{c_i} c_{1234} - \frac{c_i}{s_i} s_{1234} \, \Big) r + 2m \, \frac{c_i}{s_i} s_{1234} \Big] d \phi, \\ 
A^4_{3D} &=& \frac{2 m c_4 s_4 (r^2-2 m r +a^2) \cos{\theta}}{(r^2-2 m r +a^2 \cos^2{\theta})} d \phi.
\eea
The variables $\widetilde{\zeta}_i$ and $\zeta^4$ take the form
\bea 
\widetilde{\zeta}_i &=& \frac{m}{W^2} \Big[ 4m \Big( \frac{s_i}{c_i} \, c_{1234} \, r + (2m-r) \frac{c_i}{s_i} s_{1234} \Big)
\Big(r\, c_{1234} + (2m-r) \, s_{1234} \Big) \no \\
&& + \big( r(r-2m)+a^2\cos^2\theta \big) \Big( 4m\, (1+2s_i^2) \frac{c_{1234} \, s_{1234}}{c_i \, s_i} + 2c_i s_i \big(r-2m\, \Xi_i \big) \Big) \Big]
\eea
where
\be 
\Xi_i = \frac{2 \, (s_{1234})^2}{s_i^2} + \sum_{\substack{I<J, \\ I,J \neq i}} s_I^2 s_J^2,
\ee
and 
\bea 
\zeta^4 &=& \frac{2 a \, m \cos\theta}{W^2} \Bigg[ c_{123} \, s_4 \Big( (2 m - r) r - a^2 \cos^2\theta - 2 m r c_4^2 \Big) \no \\
&& + c_4 \, s_{123} \Big( a^2 \cos^2\theta + r(r - 2 m) + 2 m (r -2 m) s_4^2 \Big) \Bigg].
\eea

The three dilatons and three axions can be written in the following compact form
\be 
e^{\varphi_i} = \frac{r^2 + a^2 \cos^2{\theta} + g_i}{W}, \qquad \chi_i = - \frac{f_i}{{r^2 + a^2 \cos^2{\theta} + g_i}}, \label{sca1}
\ee
where 
\bea 
g_1 &=& 2 m r \, (s_1^2 + s_4^2) + 4 m^2 \, s_1^2 s_4^2, \nonumber \\
g_2 &=& 2 m r \, (s_2^2 + s_4^2) + 4 m^2 \, s_2^2 s_4^2, \nonumber \\
g_3 &=& 2 m r \, (s_3^2 + s_4^2) + 4 m^2 \, s_3^2 s_4^2, \label{sca2}
\eea
and 
\bea 
f_1 &=& 2 a m \, \cos{\theta} \, (s_{14} \, c_{23} - c_{14} \, s_{23} ), \nonumber \\
f_2 &=& 2 a m \, \cos{\theta} \, (s_{24} \, c_{13} - c_{24} \, s_{13} ), \nonumber\\
f_3 &=& 2 a m \, \cos{\theta} \, (s_{34} \, c_{12} - c_{34} \, s_{12} ). \label{sca3}
\eea

\subsection{Supersymmetric limit} 
\label{CC_bps}

The BPS limit is obtained by scaling the mass and the charge parameters as $m \to 0$ and $\gamma_I \to \infty$ for all $I$, in such a way that the charges $q_i$ and $p^0$ remain fixed:
\bea \label{q_bps}
e^{\gamma_i} &=& \sqrt{\frac{q_i}{2m}}, \quad i =1,2,3, \\
e^{\gamma_4} &=& \sqrt{\frac{2 p^0}{m}}.
\eea
Note that we choose a slightly different normalization for the parameter $\gamma_4$ relative to $\gamma_1, \gamma_2, \gamma_3$. This choice is intentional: in later sections we adopt the conventions of the five-dimensional saddle used in \cite{Adhikari:2024zif}, and the modified scaling of $\gamma_4$ above is precisely what is needed for our four-dimensional charges to match those used there. In what follows, for ease of writing longer expressions, we introduce the following shorthand notation:
\bea 
C_0 &=& \frac{q_1 q_2 q_3 + 4 q_2 q_3 p^0 + 4 q_1 q_3 p^0 + 4 q_1 q_2 p^0}{4}, \\
C_1 &=& \frac{q_1 q_2 q_3 - 4 q_2 q_3 p^0 + 4 q_1 q_3 p^0 + 4 q_1 q_2 p^0}{64}, \\
C_2 &=& \frac{q_1 q_2 q_3 + 4 q_2 q_3 p^0 - 4 q_1 q_3 p^0 + 4 q_1 q_2 p^0}{64}, \\
C_3 &=& \frac{q_1 q_2 q_3 + 4 q_2 q_3 p^0 + 4 q_1 q_3 p^0 - 4 q_1 q_2 p^0 }{64}, \\
C_4 &=& q_1 q_2 q_3 - 4 q_2 q_3 p^0 - 4 q_1 q_3 p^0 - 4 q_1 q_2 p^0.
\eea 

In taking this limit, we leave the rotation parameter $a$ unchanged. The resulting solution is rotating, supersymmetric, but non-extremal, and can be written as
\be
ds^2_\mathrm{BPS} = - \frac{r^2 + a^2 \cos^2{\theta}}{W} (dt + \omega_3)^2 +  \frac{W}{r^2 + a^2 \cos^2{\theta}} ds^2_{base},
\ee
with
\be
ds^2_\mathrm{base} = \frac{r^2 + a^2 \cos^2{\theta}}{r^2 + a^2} dr^2 + (r^2 + a^2 \cos^2{\theta}) \, d\theta^2 + (r^2 + a^2) \sin^2{\theta} \, d\phi^2. 
\ee
The three-dimensional base metric is flat space, written in Boyer-Lindquist coordinates inherited from the non-extremal solution. The function $W$ and the one-form $\omega_3$ are given by
\bea  
W^2 &=&  (r^2 + a^2 \cos^2 \theta)^2 +  \frac{1}{64 q_1 q_2 q_3 p^0} \, \left(2 C_0 r + q_1 q_2 q_3 p^0 \right)^2 \no \\
&& + \, (r^2 + a^2 \cos^2 \theta) \left[ \Gamma_1 \, r + \Gamma_2 - \frac{C_0^2}{16 q_1 q_2 q_3 p^0} \right], \\
\omega_3 &=& \frac{a\sin^2{\theta} \, (q_1 q_2 q_3 p^0 + 2 \, C_0 \, r)}{8 \sqrt{q_1 q_2 q_3 p^0} \, (r^2 + a^2 \cos^2{\theta})} d\phi, 
\eea
where 
\bea 
\Gamma_1 &=& \frac{(q_1+q_2+q_3+4p^0)}{4}, \\
\Gamma_2 &=& \frac{q_1 q_2 + q_1 q_3 + q_2 q_3 + 4p^0(q_1+q_2+q_3)}{16}. 
\eea

The four-dimensional gauge field components $\zeta^4$ and $\widetilde{\zeta}_i$ are given by
\be
\zeta^4 = - \frac{a \cos{\theta}}{16 W^2 \, \sqrt{q_1 q_2 q_3 p^0}}   \Big[ (r^2 + a^2 \cos^2{\theta})\, C_4 - p^0 \Big( 4 C_0 r + 2 q_1 q_2 q_3 p^0 \Big)  \Big]
\ee
and 
\bea
\widetilde{\zeta}_1 &=& \frac{1}{W^2} \Bigg\{  \left(r^2 + a^2 \cos^2 \theta\right) \left[ \frac{q_1}{4} \left(r + \frac{q_2 + q_3}{4} + p^0 \right) - \frac{C_1 C_0}{q_1 q_2 q_3 p^0} \right] \no \\
&& + \frac{1}{64q_1 q_2 q_3 p^0} \, \left(32 \, C_1 r + q_1 q_2 q_3 p^0 \right) \left(2 C_0 r + q_1 q_2 q_3 p^0 \right) \Bigg\}, \\
\widetilde{\zeta}_2 &=& \frac{1}{W^2} \Bigg\{  \left(r^2 + a^2 \cos^2 \theta\right) \left[ \frac{q_2}{4} \left(r + \frac{q_1 + q_3}{4} + p^0 \right) - \frac{C_2 C_0}{q_1 q_2 q_3 p^0} \right] \no \\
&& + \frac{1}{64q_1 q_2 q_3 p^0} \, \left(32 \, C_2 r + q_1 q_2 q_3 p^0 \right) \left(2 C_0 r + q_1 q_2 q_3 p^0 \right) \Bigg\}, \\
\widetilde{\zeta}_3 &=& \frac{1}{W^2} \Bigg\{  \left(r^2 + a^2 \cos^2 \theta\right) \left[ \frac{q_3}{4} \left(r + \frac{q_1 + q_2}{4} + p^0 \right) - \frac{C_3 C_0}{q_1 q_2 q_3 p^0} \right] \no \\
&& + \frac{1}{64q_1 q_2 q_3 p^0} \, \left(32 \, C_3 r + q_1 q_2 q_3 p^0 \right) \left(2 C_0 r + q_1 q_2 q_3 p^0 \right) \Bigg\}.
\eea
The corresponding three-dimensional gauge fields are given by
\bea
(\widetilde{A}_i)_{3D} &=& -\frac{a \left( 64 r \, C_i + 2 q_1 q_2 q_3 p^0 \right) \sin^2\theta}{16 \sqrt{q_1 q_2 q_3 p^0} \, (r^2 + a^2 \cos^2\theta)} \, d\phi, \\
A^4_{3D} &=& - \frac{p^0 \, (r^2 + a^2) \cos\theta}{r^2 + a^2 \cos^2\theta} \, d\phi.
\eea
Written in terms of the black hole charges, the axions take the form
\bea 
\chi_1 &=&- \frac{a \, \cos\theta}{4 \sqrt{q_1 q_2 q_3 p^0}} \, \frac{ - q_1 q_2 q_3 - 4 q_2 q_3 p^0 + 4 q_1 q_3 p^0 + 4 q_1 q_2 p^0}
{4r^2 + (q_1 + 4 p^0) r + q_1 p^0 + 4 a^2 \cos^2\theta }, \\
\chi_2 &=& -\frac{a \, \cos\theta}{4 \sqrt{q_1 q_2 q_3 p^0}} \frac{ - q_1 q_2 q_3 + 4 q_2 q_3 p^0 - 4 q_1 q_3 p^0 + 4 q_1 q_2 p^0 }{4r^2 + (q_2 + 4 p^0) r + q_2 p^0 + 4 a^2 \cos^2\theta}, \\
\chi_3 &=& -\frac{a \, \cos\theta}{4 \sqrt{q_1 q_2 q_3 p^0}} \, \frac{- q_1 q_2 q_3 + 4 q_2 q_3 p^0 + 4q_1 q_3 p^0 - 4 q_1 q_2 p^0 }{4r^2 + (q_3 + 4 p^0) r + q_3 p^0 + 4 a^2 \cos^2\theta}.
\eea
The dilatons are given by
\be 
e^{\varphi_i} = \frac{r^2 + a^2 \cos^2{\theta} + g_i}{W},
\ee
where
\bea 
g_1 = \frac{1}{4} [(q_1 + 4p^0) r + q_1 p^0], \\
g_2 = \frac{1}{4} [(q_2 + 4p^0) r + q_2 p^0], \\
g_3 = \frac{1}{4} [(q_3 + 4p^0) r + q_3 p^0].
\eea
Since the solution above is non-extremal, it describes a finite-temperature, four-charge configuration. Its Euclidean continuation furnishes the index saddle for the four-charge black hole.

\subsection{The Bates-Denef form of the solution}
\label{sec:batesdenef}
The above solution being supersymmetric can be written in a canonical form \cite{Behrndt:1997ny, Bates:2003vx}. We call this the Bates-Denef form. Before specializing to our case, it is worth briefly summarizing the general structure of this class of solutions. For this, we adopt the conventions of \cite{Adhikari:2024zif} below, with \cite{Boruch:2023gfn, Chen:2024gmc, Boruch:2025qdq, Boruch:2025biv, Boruch:2025sie} offering an alternative recent treatment.
For this class of solutions, the four-dimensional metric takes the form
\be
ds^2 = - e^{2g} (dt + \omega)^2 + e^{-2g} d s^2_{\mathrm{base}}.
\ee
where $d s^2_{\mathrm{base}}$ is the metric on the three-dimensional flat base space and $\omega$ is a one-form on the three-dimensional base space. At this stage it is useful to introduce rescaled scalars
$Y^M$. We define,
\be
Y^M = e^{-g} \bar{h} X^M,
\ee
where $h$ is a position-dependent phase appearing throughout this construction.
Homogeneity of the prepotential, 
\be
G(Y):= F(X (Y)),
\ee
then gives 
\be
G_M := \frac{\partial}{\partial Y^M} G(Y) = e^{-g} \bar{h} F_M.
\ee
Supersymmetry, encoded in the equations of motion of supergravity, fixes  $\{Y^M - \bar{Y}^M, G_M - \bar{G}_M\}$ in terms of a set of harmonic functions $\{H^M, H_M\}$ through the generalized stabilization equations,
\bea
Y^M - \bar{Y}^M &=& i H^M, \label{GSE-1} \\
G_M - \bar{G}_M &=&  i H_M. \label{GSE-2}
\eea
Once these are solved, the full spatial profile of $Y^M$ is fixed in terms of the harmonic functions.
A broad family of half-BPS solutions of the STU model can be expressed directly in terms of these harmonic functions: the metric function is given by
\be
e^{-2g} = \Sigma = -i (Y^M \bar G_M - \bar{Y}^M G_M), 
\ee
which satisfies
\bea
\Sigma^2 &=& - \left(\sum_{M=0}^{3}H^M H_M\right)^2 + \sum_{I, J, K, J', K' = 1}^{3} C_{IJK}H^J H^K C^{IJ'K'}H_{J'}H_{K'} + 4 H^0 H_1 H_2 H_3 
\nonumber  \\ 
& & 
- 4 H_0 H^1 H^2 H^3.
\eea
The physical scalar fields are given by
\be
\chi^I = \frac{2 H^I H_I - \sum_{M=0}^{3}(H^M H_M)}{\sum_{J, K = 1}^{3}C_{IJK}H^J H^K + 2 H^0 H_I},
\ee
and
\be
\exp[-\varphi_I] = \frac{\Sigma}{\sum_{J, K = 1}^{3}C_{IJK}H^J H^K + 2 H^0 H_I}.
\ee
The gauge fields take the form
\be
\frac{1}{\sqrt{2}} A^M= - \partial_{H_M}(\log \Sigma) (d t + \omega) + A_3^M, \qquad d A_3^M = - \star_3 d  H^M. \label{AM-formula}
\ee
while the one-form $\omega$ is fixed by 
\be
\star_3 d \omega = H^M d  H_M- H_M d H^M.
\ee

For the four-charge solution introduced in the previous subsection, the corresponding harmonic functions take the two-centered form
\be 
H = h + \frac{\gamma_N}{r_N} + \frac{\gamma_S}{r_S},
\ee
with the magnetic components of the north-pole charge given by
\begin{align} \label{gamma_N_up}
	\gamma_N^0 &=
	\frac{p^0}{2\sqrt{2}}, &
	\gamma_N^1 &=
	\frac{i}{4\sqrt{2}} \sqrt{\frac{q_2 q_3 p^0}{q_1}}, \\
	\gamma_N^2 &= 
	\frac{i}{4\sqrt{2}} \sqrt{\frac{q_1 q_3 p^0}{q_2}}, &
	\gamma_N^3 &= 
	\frac{i}{4\sqrt{2}} \sqrt{\frac{q_1 q_2 p^0}{q_3}},
\end{align}
and its electric components given by
\begin{align}
	\gamma_{N0} &= 
	- \frac{i}{16\sqrt{2}} \sqrt{\frac{q_1 q_2 q_3}{p^0}} , &
	\gamma_{N1} &= 
	\frac{1}{8 \sqrt{2}} \, q_1, \\
	\gamma_{N2} &= 
	\frac{1}{8 \sqrt{2}} \, q_2, &
	\gamma_{N3} &= 
	\frac{1}{8 \sqrt{2}} \, q_3. \label{gamma_N_down}
\end{align}
The south-pole charges are fixed in terms of the north-pole charges, 
\begin{align}
	\gamma_S^M &= (\gamma_N^M)^\star, &
	\gamma_{SM} &= (\gamma_{NM})^\star,
\end{align}
and the two pole distances are given as
\begin{align}
	r_N &= r - i a \cos \theta, &
	r_S &= r + i a \cos \theta.
\end{align}
The constants appearing in the harmonic functions take values
\begin{align}
	h^0 &= \frac{1}{\sqrt{2}}, & h^1&=0,&  h^2&=0, & h^3&=0, \\
	h_0 &=0, & h_1&= \frac{1}{\sqrt{2}}, & h_2&= \frac{1}{\sqrt{2}}, & h_3&= \frac{1}{\sqrt{2}}. 
\end{align}
Matching the vectors between the two presentations requires some care, since the two parametrizations differ in their asymptotic behavior: all fields in the Chow-Comp\`ere form vanish at infinity, whereas this is not automatically the case for the fields constructed from the harmonic functions above. We have verified that the two descriptions (metric, scalars, vectors) agree once the latter are shifted by the appropriate constants.

\subsection{Another example of the  new form of attraction} 
\label{sec:newattractor}

The gravitational saddles relevant to the black hole index are generically multi-centered. Before specializing to the two-centered case of interest, we recall how the single-centered (``old'') attractor mechanism fixes the near-horizon scalars of an ordinary extremal black hole. The same algebra will reappear, applied separately at each pole, once we move to the two-centered case.

Consider the spherically symmetric, asymptotically flat extremal black hole with charges $\{P^M, Q_M\}$. Sourced by these charges at the origin, the harmonic functions take the single-pole form
\bea
H^M &=& \frac{P^M}{r} + \mathcal{O}(1), \\
H_M &=& \frac{Q_M}{r} + \mathcal{O}(1),
\eea
with $r$ the radial coordinate on the three-dimensional base and the black hole horizon located at $r = 0$. Near the horizon, the generalized stabilization equations \eqref{GSE-1}--\eqref{GSE-2} force the scalars to develop matching simple poles.
\begin{align}
	Y^M &= \frac{Y^M_*}{r} + \mathcal{O}(1), &
	G_M &= \frac{G_{M*}}{r} + \mathcal{O}(1), \\
	\bar{Y}^M &= \frac{\bar{Y}^M_*}{r} + \mathcal{O}(1), &
	\bar{G}_M &= \frac{\bar{G}_{M*}}{r} + \mathcal{O}(1),
\end{align}
The generalized stabilization equations, in the limit $r \to 0$, then uniquely fixes the residues $\{Y^M_*, \bar{Y}^M_*, G_{M*}, \bar{G}_{M*}\}$. These are called the attractor equations
\bea
Y^M_* - \bar{Y}^M_* &=& i P^M, \label{attractor-1} \\
G_{M*} - \bar{G}_{M*} &=& i Q_M. \label{attractor-2}
\eea

The two-centered solutions of actual interest replace this single pole by two: the harmonic functions are instead taken to have the form
\be
H = h + \frac{\gamma_N}{r_N} + \frac{\gamma_S}{r_S},
\ee
with $r_N$ and $r_S$ the distances to the two poles in the three-dimensional base (see Section \ref{sec:batesdenef} for their explicit locations in our case). Rather than sitting entirely at one point, the black hole's charge is now split between the two poles \cite{Boruch:2023gfn},
\bea \label{egammasum_1}
\gamma_N^M + \gamma_S^M &=& P^M, \\
\gamma_{NM} + \gamma_{SM} &=& Q_M. \label{egammasum_2}
\eea
The split is fixed by requiring each pole, taken on its own, to reproduce a version of the single-centered attractor problem above. More precisely, we require the four-dimensional metric be  flat space  at the two poles. This requirements translates into the form of the scalars as $r_N \to$ \cite{Boruch:2023gfn}, 
\begin{align}
	Y^M &= \frac{i\gamma_N^M}{r_N} + \mathcal{O}(1), &
	G_M &= \frac{i\gamma_{NM}}{r_N} + \mathcal{O}(1), \\
	\bar{Y}^M &= \mathcal{O}(1), &
	\bar{G}_M &= \mathcal{O}(1),
\end{align}
while as $r_S \to 0$,
\begin{align}
	Y^M &= \mathcal{O}(1), &
	G_M &= \mathcal{O}(1), \\
	\bar{Y}^M &= -\frac{i\gamma_S^M}{r_S} + \mathcal{O}(1), &
	\bar{G}_M &= -\frac{i\gamma_{SM}}{r_S} + \mathcal{O}(1),
\end{align}
so that the generalized stabilization equations \eqref{GSE-1}--\eqref{GSE-2} hold at both poles separately. Comparing these residues against the attractor equations \eqref{attractor-1}--\eqref{attractor-2}  combined with the sum rule \eqref{egammasum_1}--\eqref{egammasum_2} gives
\be
\{\gamma_N^M, \gamma_{NM}, \gamma_S^M, \gamma_{SM}\} = \{-iY^M_*, -iG_{M*}, i\bar{Y}^M_*, i\bar{G}_{M*}\},
\ee
together with $\gamma_N = (\gamma_S)^\star$.

The four-charge black hole  of Section \ref{CC_bps} carries charges\footnote{These factors of $\sqrt{2}$ trace back to a mismatch in how the gauge kinetic terms are normalised. The supergravity literature conventionally fixes the kinetic term for a vector to be $-\frac{1}{2} F_{\mu \nu} F^{\mu \nu}$ when the Einstein-Hilbert term has unit coefficient (see e.g.\ \cite{Mohaupt:2000mj, LopesCardoso:2000qm}), whereas we instead use the more common normalisation $-\frac{1}{4} F_{\mu \nu} F^{\mu \nu}$, as in eq.~\eqref{tilde123Lagrangian}. The rescaling of the gauge fields needed to pass between the two conventions is exactly what produces the $\sqrt{2}$ in eq.~\eqref{AM-formula}, and hence in the charges above.},
\begin{align}
	P^0 &= \frac{1}{\sqrt{2}}  p^0\, , &  Q_0 &= 0& 
	P^1 &= 0 \, , & Q_1 &= \frac{1}{4\sqrt{2}}q_1\, ,& \\
	P^2 &= 0\, , & Q_2 &= \frac{1}{4\sqrt{2}}q_2\, ,&
	P^3 &= 0\, , & Q_3 &= \frac{1}{4\sqrt{2}}q_3\, .& 
\end{align}
Solving the attractor equations \eqref{attractor-1}--\eqref{attractor-2} for these charges, and splitting the result between the two poles as in \eqref{egammasum_1}--\eqref{egammasum_2}, fixes $\gamma_N$ and $\gamma_S$ to be exactly \eqref{gamma_N_up}--\eqref{gamma_N_down}. This confirms that the index saddle constructed above, as the BPS limit of the non-extremal four-charge black hole, satisfies the new attractor mechanism.

\section{The index saddle for the D1-D5-P black string}
\label{sec:D1D5saddle}

Having constructed the index saddle for the four-charge black hole in four dimensions, a saddle that exhibits the new form of attraction, we now turn to the higher-dimensional uplift of this solution. Our goal is to obtain the gravitational index saddle for the D1-D5 black string, the six-dimensional object whose near-horizon geometry is $AdS_3 \times S^3$. We proceed in two steps. The overall logic is summarised in Figure~\ref*{fig:roadmap}. First, in Sections~\ref{sec:typeIIBuplift}--\ref{sec:harmonicfunctions}, we review the Type IIB uplift of five-dimensional supersymmetric solutions and recall the Anupam-Chowdhury-Sen (ACS) saddle that computes the supersymmetric index of the BMPV black hole, writing it in terms of harmonic functions on a three-dimensional flat base that properly include the rotation parameter $J_{3L}$. Second, in Sections~\ref{sec:4d5dconnection}--\ref{sec:indexsaddleD1D5}, we relate these harmonic functions to the attractor charges of the four-dimensional index saddle of Section~\ref{sec:4charge} via the 4D-5D connection, decompactifying the M-theory circle to recover the asymptotically flat ACS solution; since the four-charge black hole of Section~\ref{sec:4charge} does not carry $J_{3L}$, this matching is established explicitly only in the case $J_{3L}=0$. We then uplift the general ACS solution to six dimensions, following the procedure of Section~\ref{sec:typeIIBuplift}, to obtain the index saddle for the D1-D5 black string. This construction sets the stage for Section~\ref{sec:decoupling}, where we take the $AdS_3 \times S^3$ decoupling limit of the resulting geometry. The order of the two steps is dictated by the nature of the matching. The 4D-5D connection is
established at the level of the harmonic functions, it can therefore only be stated once the harmonic functions
are in place.

\subsection{Type IIB uplift of the 5D supersymmetric solutions}
\label{sec:typeIIBuplift}

The supersymmetric structure of the ACS saddle is most transparent when the solution
is lifted to eleven dimensions and interpreted in terms of intersecting M2-branes.
Bena and Warner~\cite{Bena:2005va, Bena:2007kg} showed that solutions preserving the same
supersymmetries as three-charge black holes and black rings can be written in a
general form, in terms of one-forms and functions defined on a four-dimensional
hyper-K\"ahler base space. Their construction is simplest and most symmetric when
phrased in M-theory, with three stacks of M2-branes wrapping orthogonal two-tori
inside a six-torus with coordinates $(z_1, \ldots, z_6)$, intersecting according to
the standard pattern M2(12)--M2(34)--M2(56). This M-theory frame is also the natural starting point for reaching the D1-D5
system; a chain of dualities and dimensional
reductions maps this M2-M2-M2 configuration onto the D1-D5-P system of Type~IIB
string theory on $T^4 \times S^1$.

The eleven-dimensional metric takes the symmetric form
\be
ds_{11}^2 = ds_5^2+ ds^2_{\mathrm{T}^6},
\ee
where $ds^2_{\mathrm{T}^6}$ is the metric on the six-torus,
\be \label{T6_metric_BW}
ds^2_{\mathrm{T}^6} = (Z_2 Z_3 Z_1^{-2})^{\frac{1}{3}}(dz_1^2+dz_2^2)
+(Z_1Z_3Z_2^{-2})^{\frac{1}{3}}(dz_3^2+dz_4^2)+(Z_1Z_2Z_3^{-2})^{\frac{1}{3}}(dz_5^2+dz_6^2),
\ee
and $ds_5^2$ is the metric on the five-dimensional transverse spacetime,
\be \label{ds5_BW}
ds_5^2=-(Z_1Z_2Z_3)^{-\frac{2}{3}}(dt+k)^2+(Z_1Z_2Z_3)^{\frac{1}{3}}h_{mn}dx^m dx^n,
\ee
where $h_{mn}$ is the metric on a four-dimensional hyper-K\"{a}hler base space. The
three functions $Z_I$, $I=1,2,3$, are sourced by the three pairs of M2-brane stacks and
will become, after the duality chain, the
harmonic functions associated with the D1, D5, and momentum charges.

The M-theory three-form potential $\mathcal{A}$ for this class of solutions can be
written in terms of three one-form potentials $A^{(I)}$ on the five-dimensional
spacetime, one for each brane pair,
\be
\mathcal{A}=A^{(1)}\wedge dz_1\wedge dz_2+A^{(2)}\wedge dz_3\wedge dz_4+A^{(3)}\wedge dz_5\wedge dz_6,
\ee
which in turn take the form,
\be
A^{(I)}  = -\frac{(dt+k)}{Z_I}+\omega_I , \label{A_I}
\ee
where $k$ and $\omega_I$ are one-forms on the four-dimensional base space while $Z_I$
are functions on the base space. These functions and one-forms are determined by the
BW equations~\cite{Bena:2005va}:
\begin{eqnarray}
	d\omega_{I} &=& \star  d\omega_I, \label{BW_EQ_1}\\
	dk + \star  dk &=& Z_I d\omega_I, \label{BW_EQ_2} \\
	\nabla^2 Z_I &=& \frac{1}{2}|\epsilon_{IJK} | \star  (d\omega_J \wedge d\omega_K),\label{BW_EQ_3}
\end{eqnarray}
where the Hodge star is with respect to the four-dimensional base metric $h_{mn}$.

The eleven-dimensional metric and three-form potential above solve the equations of motion following from the eleven-dimensional supergravity action
\be \label{lag_11}
I_{11} = \frac{1}{16\pi G_{11}}\int \left( R_{11}\star_{11}1 - \frac{1}{2}\mathcal{F}\wedge\star_{11}\mathcal{F}-\frac{1}{6}\mathcal{F}\wedge\mathcal{F}\wedge\mathcal{A}\right),
\ee
where $\mathcal{F}=d\mathcal{A}$. More generally, before imposing any supersymmetry condition, reducing this action on $T^6$ via the ansatz
\bea \label{ansatz_11to5}
ds_{11}^2 &=& ds_5^2 + h^1\!\left(dz_1^2+dz_2^2\right)+h^2\!\left(dz_3^2+dz_4^2\right)+h^3\!\left(dz_5^2+dz_6^2\right), \\
\mathcal{A} &=& A^1\wedge dz_1\wedge dz_2 + A^2\wedge dz_3\wedge dz_4 + A^3\wedge dz_5\wedge dz_6,
\eea
with nothing depending on the $T^6$ coordinates, gives five-dimensional $U(1)^3$ supergravity, in which $ds_5^2$, $A^I$, and $h^I$ are an arbitrary Einstein frame five-dimensional metric, three vectors, and three scalars subject to 
\be
h^1 h^2 h^3=1.
\ee 
The Bena-Warner solution~\eqref{T6_metric_BW}--\eqref{ds5_BW} is the special, supersymmetric case of this reduction with
\begin{align}
	h^1&=(Z_2Z_3Z_1^{-2})^{1/3}, & h^2&=(Z_1Z_3Z_2^{-2})^{1/3}, & h^3&=(Z_1Z_2Z_3^{-2})^{1/3}, 
\end{align}
and  $A^I = A^{(I)}$, consistent with the $T^6$ metric~\eqref{T6_metric_BW}. The five-dimensional theory obtained this way is precisely $\mathcal{N}=2$ supergravity coupled to two vector multiplets: a graviphoton together with the two vectors from the vector multiplets account for the three gauge fields $A^I$, and the two independent scalars surviving the constraint $h^1h^2h^3=1$ parametrize the moduli space with metric $G_{IJ}=\tfrac12(h^I)^{-2}\delta_{IJ}$. This theory is also known as the 5D STU model or $U(1)^3$ supergravity.

The eleven-dimensional origin makes the brane content of the 5D charges manifest: examining $\mathcal{F}$, the electric charge $Q_I$ coupling to $F^I=dA^I$ is sourced by M2-branes wrapping the corresponding $T^2\subset T^6$, e.g.\ $F^1$ by M2-branes wrapping the $12$-cycle, and is quantized in terms of the M2-brane wrapping number.

We can map this five-dimensional theory directly to the duality frame in which the M2 brane system becomes the D1-D5 system with momentum running along their common direction. Reducing the eleven-dimensional solution on the $z_6$ circle gives a solution of Type IIA supergravity; performing T-dualities along $z_5, z_4, z_3$ then gives a solution of Type IIB supergravity with the string frame metric
\be \label{IIB_metric}
ds^2_{\mathrm{IIB}} = (h^3)^{1/2}ds_5^2 + (h^3)^{-3/2}(dy+A^3)^2 + h^1(h^3)^{1/2}dz_{(4)}^2,
\ee
where the $T^4 \times S^1$ is parametrized by  $z_{(4)}\equiv(z_1,z_2,z_3,z_4)$ and $y\equiv z_5$. The circle parametrized by $y$, which we will refer to as the Kaluza-Klein circle, is in general nontrivially fibered over the five-dimensional spacetime. The remaining nonzero Type IIB fields are (see, e.g., \cite{Emparan:2006mm})
\be \label{IIB_dilaton_RR}
e^{2\Phi} = \frac{h^1}{h^2}, \qquad
F_{(3)} = (h^1)^{-2}\star_5 F^1 + F^2 \wedge (dy+A^3),
\ee
with $\Phi$ the dilaton and $F_{(3)}$ the Ramond-Ramond three-form field strength. Equations~\eqref{IIB_metric}--\eqref{IIB_dilaton_RR} uplift any solution of five-dimensional $U(1)^3$ supergravity to Type IIB string theory. Examining the RR three-form, the electric charges coupling to $F^1$ and $F^2$ are sourced by D5-branes wrapped on $T^4 \times S^1$ and D1-branes wrapped around the $y$-circle, respectively, while the appearance of $A^3$ in the metric shows that $F^3$ is sourced by momentum $P$ around the Kaluza-Klein circle. The six-dimensional D1-D5 black string is obtained by simply suppressing the four $T^4$ directions.

\subsection{The ACS saddle for the BMPV black hole}
\label{sec:ACSsaddle}

We now turn to a specific solution of the five-dimensional $U(1)^3$ supergravity introduced above. Anupam, Chowdhury, and Sen~\cite{Anupam:2023yns} constructed a five-parameter solution of this theory that saturates the BPS bound while remaining at finite temperature -- precisely the kind of supersymmetric, non-extremal saddle that the index program described in the Introduction requires. We refer to this solution as the ACS saddle. It is the Euclidean continuation of an asymptotically flat black hole, carrying three $U(1)$ charges $Q_1, Q_2, Q_3$ and two independent angular momenta $J_{3L}, J_{3R}$. The metric is written in $(t,\rho,\theta,\phi,\psi)$ coordinates with the standard ranges (these coordinates should not be confused with those of Section~\ref{sec:4charge}), and takes the compact form~\cite{Adhikari:2024zif}:
\begin{equation}
	d s^2= -f^2(d t + k_\phi d \phi + k_\psi d \psi)^2+f^{-1} d s^2_{\mathrm{4d-base}},
	\label{5dmetric}
\end{equation}
with
\bea
f  &=& \frac{\cos (2 \theta ) J_{3L} J_{3R}+2 \rho^2 Q_S}
{2 \, \Delta^{1/3}  \, Q_S}, \label{ACS_f} \\
k_\phi &=& \frac{Q_S\, \sin^2 \theta\,[4 \jr Q_T+ 2(\jr+\jl) Q_S \, \rho^2 +\jl \jr \, (\jr+\jl ) \cos(2\theta)]}{(2Q_S\, \rho^2+\jl\jr\,\cos(2\theta))^2},
\nonumber \\	\label{omegaphi}\\
k_\psi &=& \frac{Q_S\, \cos^2 \theta\,[4 \jr Q_T+ 2(\jr-\jl) Q_S \, \rho^2 +\jl \jr \, (\jr-\jl ) \cos(2\theta)]}{(2Q_S\, \rho^2+\jl\jr\,\cos(2\theta))^2}.
\nonumber \\
\label{omegapsi}
\eea
Here,
\bea
Q_S &=& Q_1 Q_2 + Q_2 Q_3 + Q_3 Q_1, \label{eq:Q_S} \\
Q_T &=& Q_1 Q_2 Q_3, \\ 
\Delta& =& \frac{1}{8Q_S^{\, 3}}\prod\limits_{I=1}^3[\cos(2\theta)J_{3L}J_{3R}+2(Q_I + \rho^2)Q_S].
\eea
The  four-dimensional base metric $d s^2_{\mathrm{4d-base}}$ is Ricci flat and is given as,
\be \label{4d_base}
d s^2_{\mathrm{4d-base}}= g_{\rho \rho} \, d \rho^2 + g_{\theta \theta} \, d \theta^2 + g_{\phi \phi} \, d \phi^2+ g_{\psi \psi} \, d \psi^2 + 2 g_{\phi \psi} \, d \phi d \psi,
\ee
where
\bea
g_{\rho \rho} &=& \rho^2 \, \frac{\rho^2 +  c \, d \, \cos 2 \theta}{\rho^4 +  d^2 \, (1-c^2)}\, ,\\
g_{\theta \theta} &=&\rho^2+  c \, d \, \cos 2\theta \,  \, , \\
g_{\phi \phi} &=&(\rho^4 + 2  \, c  \, d \,  \rho^2 \cos^2 \theta + d^2  \sin^2 \theta + c^2  \, d^2  \cos 2 \theta) \frac{\sin^2 \theta}{\rho^2+  c\, d\, \cos 2\theta}\, ,\\
g_{\psi \psi} &=&  (\rho^4 - 2  \, c  \, d \,  \rho^2 \sin^2 \theta + d^2  \cos^2 \theta - c^2  \, d^2  \cos 2 \theta) \frac{\cos^2 \theta}{\rho^2+  c\, d\, \cos 2\theta}\, ,\\
g_{\phi \psi} &=&  \frac{d^2 \sin^2 \theta \cos^2 \theta}{\rho^2 + c \, d \, \cos 2 \theta}\,.
\eea
where we have introduced two new parameters,
\bea	
c&=& \frac{\jl}{\sqrt{4 Q_1 Q_2 Q_3}}, \\
d&=&\frac{\jr\sqrt{Q_1 Q_2 Q_3}}{Q_1 Q_2 + Q_2 Q_3 + Q_3 Q_1}.
\eea
The range of the parameters is restricted to $Q_I >0$ and $J_{3L}$ is taken to be real such that 
\be \label{ACS_D} 
D = \sqrt{4 Q_1 Q_2 Q_3 - J_{3L}^2} > 0.
\ee
$J_{3R}$ is taken to lie along the positive imaginary axis, $-i \, J_{3R} > 0$. This in turn implies that the new parameters are such that
\be \label{ACS_restrictions}
(1-c^2) > 0 \quad \text{and} \quad d^2 < 0.
\ee   
We have focused here on the metric; the corresponding vector fields and scalars of the ACS saddle can be found in~\cite{Adhikari:2024zif}.

\subsection{Harmonic functions for the ACS saddle}
\label{sec:harmonicfunctions}

Section~\ref{sec:typeIIBuplift} reviewed the result of Bena and Warner~\cite{Bena:2005va} that
any solution preserving the same supersymmetries as the five-dimensional three-charge black
hole can be written in terms of a set of functions and one-forms on a four-dimensional
hyper-K\"ahler base space, subject to the linear system~\eqref{BW_EQ_1}--\eqref{BW_EQ_3}. A
substantial further simplification occurs when this hyper-K\"ahler base is of Gibbons-Hawking
type, that is, when it admits a tri-holomorphic $U(1)$ isometry. The base is then fixed by a
single harmonic function $V$ on a flat three-dimensional space, and the Bena-Warner equations
are solved by taking the warp factors $Z_I$ and the one-forms $k$ and $\omega_I$ to be determined by eight harmonic functions $(V, K^I, L_I, M)$ with $I =1,2,3$ on this flat base. 

The ACS saddle of Section~\ref{sec:ACSsaddle}, presented there in its original
$(t, \rho,\theta,\phi,\psi)$ form, falls within this class: its four-dimensional base is of
Gibbons-Hawking form. The general ACS solution was written in terms of these eight harmonic
functions in~\cite{Adhikari:2024zif}, and we recast the ACS metric in this form here, presenting
the main points.

Concretely, the base \eqref{4d_base} can be written in Gibbons-Hawking form,
\be	\label{GibbHawk}
\diff s^2_{\mathrm{4d-base}} = V^{-1}(d z+A)^2+ V d s^2_{\mathrm{3d-base}}.
\ee
Reaching this form requires the adapted angular coordinates
\be
z = \psi - \phi, \qquad \varphi = \phi + \psi,
\ee
in terms of which the flat three-dimensional base $\diff s^2_{\mathrm{3d-base}}$ is expressed through $(\rho, \theta, \varphi)$. Matching \eqref{GibbHawk} against the explicit base \eqref{4d_base} fixes the Gibbons-Hawking potential to
\be \label{harmonicV}
V=\frac{4\,(\rho^2+c \, d\, \cos 2\theta)}{\rho^4+(1-c^2) \, d^2 \, \cos^2 2\theta} \, ,
\ee
while the connection one-form $A$ is determined by
\be \label{stardV}
\star_3 d A=d V.
\ee
Here $\star_3$ is the Hodge dual on the three-dimensional base, with orientation fixed by the convention $\epsilon_{\rho \theta \varphi} > 0$. Written out, this base metric is
\begin{align}
	\diff s^2_{\mathrm{3d-base}}=& \, \frac1{4} \, (\rho^4+(1-c^2) \, d^2 \cos^2 2\theta) \, \bigg\{\frac{\rho^2}{\rho^4+(1-c^2)\, d^2}\, \diff\rho^2+\diff \theta^2\bigg\} \nonumber\\
	&+ \frac{1}{16} \, (\rho^4+(1-c^2)\,d^2) \sin^2 2\theta\, \diff\varphi^2.
\end{align}
The remaining harmonic functions are obtained as follows. The one-form $k$ entering the five-dimensional metric splits into a fibre piece and a base piece,
\be
k=\mu(\diff z +A) +\omega_{BW}. \label{omega-1}
\ee
Its scalar coefficient $\mu$ is assembled from all eight harmonic functions,
\be
\mu= \frac{1}{6}C_{IJK} \frac{K^IK^JK^K}{V^2}+\frac{1}{2V}K^IL_I +M,
\ee
where $C_{IJK} = |\epsilon_{IJK}|$ and the three-dimensional one-form $\omega_{BW}$ is the solution of
\be
\star_3\diff\omega_{BW} = V\diff M-M\diff V+\frac12(K^I\diff L_I-L_I\diff K^I).  \label{omega-2}
\ee
The warp factor $f$ is the combination
\be
f=(Z_1Z_2Z_3)^{-1/3}
\ee
of the three electric functions $Z_I$, each quadratic in the magnetic harmonic functions,
\be
Z_I = \frac{1}{2V} C_{IJK} K^J K^K+L_I.
\ee
The three Maxwell fields of the solution read
\be
A^I =- \frac{1}{Z_I} (\diff t+k) + \frac{K^I}{V}(\diff z + A)+\xi^I + \diff t, 
\ee
where the magnetic one-forms $\xi^I$ on the base obey
\be	
\star_3\diff\xi^I=-\diff K^I.
\ee
The three scalar moduli are then
\be
h^I = (fZ_I)^{-1}.
\ee

This fixes the generic Bena-Warner dictionary; what remains is to give the eight harmonic functions for the ACS saddle itself. The saddle is two-centered, so each harmonic function $H \in \{V, K^I, L_I, M\}$ is sourced at two points,
\be
H = h + \frac{\gamma_N}{r_N} + \frac{\gamma_S}{r_S}, 
\ee
the centers sitting at $\vec y_N$ and $\vec y_S$ with $r_N$ and $r_S$ the corresponding distances on the flat base,
\bea
r_N &=& |\vec y - \vec y_N| = \sqrt{(y^1 - y^1_N)^2 + (y^2 - y^2_N)^2 + (y^3 - y^3_N)^2},   \label{north-pole-location-1} \\
r_S &=& |\vec y - \vec y_S| = \sqrt{(y^1 - y^1_S)^2 + (y^2 - y^2_S)^2 + (y^3 - y^3_S)^2}. \label{south-pole-location-1}
\eea
For the ACS saddle the two centers lie on the symmetry axis at
\begin{align}
	\vec y_N &= \{y^1_N, y^2_N,y^3_N\} =\left\{0,0, -\frac i4 \sqrt{1-c^2} \, d\right\} \label{north-pole-location}, \\
	\vec y_S &= \{y^1_S, y^2_S,y^3_S\} =\left\{ 0,0, \frac i4 \sqrt{1-c^2} \, d\right\} \label{south-pole-location}.
\end{align}
Re-expressed in the $(\rho,\theta,\varphi)$ coordinates introduced above, the two distances are
\begin{align}
	&r_N=\frac14(\rho^2+i\sqrt{1-c^2} \, d \cos 2\theta ),\\
	&r_S=\frac14(\rho^2-i\sqrt{1-c^2} \, d \cos 2\theta ).
\end{align}
Note that the parameter restrictions \eqref{ACS_restrictions} keep $r_N$ and $r_S$ real. The eight harmonic functions are
\begin{align}
	&K^I = \frac{k^I_N}{r_N}+ \frac{k^I_S}{r_S}, &
	&V = \frac{v_N}{r_N}+\frac{v_S}{r_S}, &  \label{ACS_harmonic_1}\\
	&L_I= 1+ \frac{l_{IN}}{r_N}+\frac{l_{IS}}{r_S}, & 
	&M= \frac{m_N}{r_N}+\frac{m_S}{r_S},  \label{ACS_harmonic_2}&
\end{align}
with the residues at the two poles given by
\begin{align} 
	&v_N = (v_S)^* = \frac12\bigg(1+i\frac {J_{3L}}{D}\bigg) , \label{ACS_charges_1}\\
	&k^I_N = (k^I_S)^*  = \frac i{2}\frac{Q_1 Q_2 Q_3}{Q_I  D} , \\
	&m_N = (m_S)^* = - \frac1{32} \left(J_{3L}+\frac{i}{2}\left(D - \frac{J_{3L}^2}{D}\right)\right), \\
	&l_{IN} = (l_{IS})^*  = \frac{ 1}{8} Q_I \left(1-i \frac {J_{3L}}{D}\right),\label{ACS_charges_4} 
\end{align}
with $D$ given in \eqref{ACS_D}. As noted in Section~\ref{CC_bps}, writing a 4D metric in the
Bates-Denef form guarantees the existence of Killing spinors; similarly, writing a 5D metric in the
Bena-Warner form guarantees that the solution is supersymmetric. The Killing spinors for the
general Bena-Warner class were constructed in \cite{Bena:2004de}, and those of the ACS saddle
specifically in \cite{Hegde:2023jmp}, confirming that the two-centered solution above indeed
preserves supersymmetry. With the ACS saddle now cast in harmonic-function form, with explicit
pole charges~\eqref{ACS_charges_1}--\eqref{ACS_charges_4}, we are ready to match these against
the four-dimensional attractor charges in Section~\ref{sec:4d5dconnection}.

\subsection{4D-5D connection and decompactification}
\label{sec:4d5dconnection}

In Section~\ref{sec:newattractor}, we saw that the four-dimensional, four-charge index saddle
exhibits the new attractor mechanism and can be written in terms of eight harmonic functions
$(H^M, H_M)$. In Section~\ref{sec:harmonicfunctions}, we wrote the ACS saddle, with its two
independent rotations $J_{3L}$ and $J_{3R}$, in Bena-Warner form in terms of its own eight
harmonic functions $(V, K^I, L_I, M)$. We now show that, under a suitable dictionary between
the 4D and 5D charges sourcing these harmonic functions, the two sets coincide: the
four-dimensional index saddle, uplifted along the Kaluza-Klein circle, reproduces the ACS
saddle.

The key to this identification is that the four 4D gauge fields $(A^0, A_i)$ do not all
play the same role. The three fields $A_i$ map directly onto the gauge fields $A^I$ of
the five-dimensional STU model, their charges becoming, the three
$U(1)$ charges of the ACS saddle,
\be
q_1 = Q_1, \qquad q_2 = Q_2, \qquad q_3 = Q_3.
\ee
The fourth field $A^0$ is the Kaluza-Klein vector, and its role is geometric: the uplift
proceeds along it, generating the fifth dimension and casting the four-dimensional spatial base
in Taub-NUT form, a circle fibered over $\mathbb{R}^3$. The magnetic charge $p^0$ sourcing $A^0$
is the Taub-NUT charge of this base, and requiring the uplifted solution to be asymptotically
flat fixes it to a single unit, $p^0 = 1$, so that the base is globally $\mathbb{R}^4$ rather
than the orbifold $\mathbb{R}^4/\mathbb{Z}_{p^0}$ that a higher charge would produce. Finally,
the rotation parameter $a$ of the 4D black hole is related  with the 5D angular momentum
$J_{3R}$.

Taking  this into account for the attractor charges we get the corresponding split of charges \eqref{gamma_N_up}--\eqref{gamma_N_down} between the north and the south poles as
\begin{align} 
	&\gamma_N^0 = \frac{1}{2\sqrt{2}},&
	&\gamma_{N0} = - \frac{i}{16\sqrt{2}} \sqrt{Q_1 Q_2 Q_3}, & \label{ACS_charges_de1} \\
	&\gamma_N^1 = \frac{i}{4\sqrt{2}} \sqrt{\frac{Q_2 Q_3}{Q_1}}, &
	&\gamma_{N1} = \frac{1}{8 \sqrt{2}} \, Q_1,& \\
	&\gamma_N^2 = \frac{i}{4\sqrt{2}} \sqrt{\frac{Q_1 Q_3}{Q_2}}, &
	& \gamma_{N2} = \frac{1}{8 \sqrt{2}} \, Q_2, & \\
	&\gamma_N^3 = \frac{i}{4\sqrt{2}} \sqrt{\frac{Q_1 Q_2}{Q_3}}, &
	&\gamma_{N3} = \frac{1}{8 \sqrt{2}} \, Q_3. & \label{ACS_charges_de4}
\end{align}
These charges match exactly with \eqref{ACS_charges_1}--\eqref{ACS_charges_4} of the Bena-Warner harmonic functions for $J_{3L}=0$ showing that the four dimensional index saddle can be uplifted to the five dimensional ACS saddle solution. For this matching, we also need to take into account the correct normalization in the relation between the eight new attractor harmonic functions $\{ H^M, H_M \}$ and the eight Bena-Warner harmonic functions:
\begin{align} 
	&H^0 = \frac{1}{\sqrt{2}} \, V, & \qquad  &H_0 = \sqrt{2} \, M, & \label{harmonic_4D_5D_1} \\
	&H^I = \frac{1}{\sqrt{2}} \, K^I, & \qquad  &H_I = \frac{1}{\sqrt{2}} \, L^I. &  \label{harmonic_4D_5D_2}
\end{align}
Decompactification simply corresponds to dropping the constant term in $H^0$.

We note here that extending this matching to non-zero $J_{3L}$ is only a technical complication, not a conceptual one. In the four-dimensional description, the parameter $J_{3L}$ corresponds to the electric charge of the Kaluza-Klein vector field $A^0$. It is precisely one of the charges that we set to zero in Section \ref{sec:4charge} when constructing the non-extremal solution. Matching all parameters at non-zero $J_{3L}$ therefore requires extending that solution to carry one additional charge, the electric charge of $A^0$.

Turning on this charge is not straightforward. Starting from the neutral seed and applying the relevant boost or duality transformation naively, introduces a Lorentzian NUT charge. Such a charge  renders the resulting geometry pathological: it develops closed timelike curves. A simple example exhibits this obstruction \cite{Elvang:2003mj}. Take a single Kaluza-Klein monopole in five dimensions,
\be
ds^2 = -dt^2 + H^{-1}\left[dz + q(\cos\theta + 1)\, d\phi\right]^2 + H\left(dr^2 + r^2 d\Omega_2^2\right), \qquad H = 1 + \frac{q}{r}.
\ee
Its only subtlety at $\theta = 0$ is a coordinate artifact, harmless once the patches are chosen correctly. The simplest way to attempt to turn on the desired electric charge is to boost along $z$. This mixes $t$ and $z$ in the fiber over the $S^2$, and the metric becomes
\bea \label{boostedKK}
ds^2 &=& \frac{H_\alpha}{H}\left(dz - \frac{q \sinh 2\alpha}{2 r H_\alpha}\, dt + \frac{q \cosh\alpha}{H_\alpha}(\cos\theta+1)\, d\phi \right)^2 \no \\
&& - H_\alpha^{-1}\left(dt - q \sinh\alpha (\cos\theta + 1)\, d\phi\right)^2 + H\left(dr^2 + r^2 d\Omega_2^2\right),
\eea
where
\be
H_\alpha = 1 - \frac{q \sinh^2\alpha}{r}.
\ee
The result is that what was a removable coordinate artifact before the boost becomes a genuine Lorentzian NUT charge after it. The reason is simply that the periodic identifications required by the KK-monopole fibration and those required by the boost cannot be satisfied simultaneously.

This obstruction is well understood, and is known to be removable \cite{Rasheed:1995zv, Larsen:1999pp, Compere:2010fm, Chow:2013tia, Chow:2014cca}. One introduces an extra parameter already in the seed solution, and tunes it so that the Dirac-Misner string cancels once the desired charge is turned on (in our example the Kaluza-Klein electric charge). Carrying this out for the four-charge non-extremal solution of Section \ref{sec:4charge} would require working with a considerably more complicated solution carrying further parameters. We see no genuine obstruction to completing this construction, but it would add little to what we have already learned from the $J_{3L} = 0$ case, so we do not pursue it here.

We also note that, although this complication affects the non-extremal construction, it is absent for the BPS solution itself. The corresponding BPS two-center solution can be constructed directly from the new attractor mechanism, without first passing through a non-extremal seed, and matches the five-dimensional ACS saddle under the 4D-5D connection. This calculation was recently reported by Larsen and Sharma \cite{Larsen:2026sav}.

\subsection{The index saddle for the D1-D5-P black string}
\label{sec:indexsaddleD1D5}

We now have every ingredient needed to write down the six-dimensional index saddle. The
ACS saddle has been cast in Bena-Warner form, and its eight harmonic functions have been
fixed in terms of the attractor charges of the four-dimensional index saddle through the
dictionary of Section~\ref{sec:4d5dconnection}. It remains only to insert this solution into
the Type~IIB uplift~\eqref{IIB_metric}--\eqref{IIB_dilaton_RR} of
Section~\ref{sec:typeIIBuplift} and read off the resulting ten-dimensional fields.

Carrying out the uplift for the ACS saddle, the three functions $Z_I$ become the harmonic
functions of the three charges of the system, the D5, D1, and momentum charges identified
below~\eqref{IIB_dilaton_RR}. In the frame in which the solution takes its standard D1-D5-P
form, the six-dimensional metric is
\be
ds_{6}^2=-\frac{1}{Z_3\sqrt{Z_1Z_2}}(dt+k)^2+\sqrt{Z_1Z_2}\,h_{mn}dx^m dx^n + \frac{Z_3}{\sqrt{Z_1Z_2}} (dy + A^3)^2,
\label{D1D5_metric}
\ee
with dilaton
\be
e^{2\Phi} = \frac{Z_2}{Z_1},
\label{D1D5_dilaton}
\ee
and Ramond-Ramond three-form
\be
F_{(3)} = (Z_2 Z_3 Z_1^{-2})^{-2/3}\star_5 F^1 + F^2 \wedge (dy+A^3).
\label{D1D5_F3}
\ee

This solution inherits its regularity and supersymmetry from its lower-dimensional ancestors.
Casting the ACS saddle in Bena-Warner form already guarantees the existence of Killing
spinors~\cite{Hegde:2023jmp, Adhikari:2024zif,  Boruch:2025qdq} (Section~\ref{sec:harmonicfunctions}), and the uplift
formulas~\eqref{IIB_metric}--\eqref{IIB_dilaton_RR} preserve these spinors. The six-dimensional
solution~\eqref{D1D5_metric}--\eqref{D1D5_F3} is therefore smooth wherever the ACS saddle itself
is smooth, and the periodic fermion boundary conditions that implement the $(-1)^F$ insertion
around the thermal circle descend unchanged through the uplift.

For the same reason, the on-shell action of this saddle is $\beta$-independent. Section~\ref{sec:4d5dconnection} showed, at $J_{3L}=0$, that the pole charges of the ACS saddle's harmonic
functions coincide with those fixed by the new attractor mechanism~\cite{Boruch:2023gfn} of
Section~\ref{sec:newattractor}: the scalar moduli at the centers $\vec y_{N,S}$ depend only on the
charges $Q_1, Q_2, Q_3$, not on $\beta$ or on the asymptotic moduli. For $J_{3L}\neq 0$, this same
new-attraction property of the ACS saddle's pole charges was established in
\cite{Adhikari:2024zif}. Dimensionally
reducing~\eqref{D1D5_metric}--\eqref{D1D5_F3} on the $S^1$ relates the
six-dimensional on-shell action to the five-dimensional one by an overall factor
$\mathrm{Vol}(S^1) = (2\pi R)$, fixed by the asymptotic moduli  $R$
and is independent of $\beta$. However, there is a subtlety: bringing the reduced action back to the canonical $U(1)^3$
form of Section~\ref{sec:typeIIBuplift}, requires an electromagnetic duality on one of
the field strengths, whose effect on the on-shell action we have not tracked. We  expect
the $\beta$-independence of the gravitational index for the D1-D5 black string to hold for the
general ACS saddle, but it will be nice to confirm this expectation via an explicit calculation.

The solution \eqref{D1D5_metric}--\eqref{D1D5_F3} is the gravitational index saddle for the D1-D5 black string. Its near-horizon region,
where the $AdS_3 \times S^3$ structure emerges is the subject of Section~\ref{sec:decoupling}.

\section{The $AdS_3 \times S^3$ decoupling limit}
\label{sec:decoupling}

The six-dimensional index saddle constructed in the previous section is asymptotically flat in five-dimensions; asymptotically Kaluza-Klein in six-dimensions.
To make contact with the AdS$_3$/CFT$_2$ correspondence, and ultimately with the index of the D1-D5 CFT, we must isolate the low-energy excitations of the D1-D5 bound state
and decouple them from the flat asymptotic region. On the gravity side this is the
decoupling limit, in which the geometry develops a long throat of the form
$AdS_3 \times S^3$, fibred trivially over the spectator $T^4$ and weakly coupled to the flat
exterior. The limit can be reached by making the radius $R$ of the common D1-D5 circle the largest
scale in the problem. This limiting procedure is standard in the fuzzball literature (see, for example, \cite{Chakrabarty:2015foa}), and we adopt it here.

The ten-dimensional Newton's
constant for Type IIB theory is
\be
G_{10} = 8 \pi^6 g^2 \alpha'{}^4, 
\ee
where $g$ is the string coupling and $\alpha'$ is related to the string length $l_s$ as $\alpha' = \ell_s^2$. 
Denoting the asymptotic length of the $S^1$ by $2\pi R$ and the volume of the $T^4$ by
$(2\pi \ell)^4$, the quantized integral charges are \cite{Maldacena:1996ky}
\bea
Q_1 &=&  g \alpha' N_5, \\
Q_2 &=& \frac{g \alpha'{}^3}{\ell^4} N_1, \\
Q_3 &=& \frac{g^2 \alpha'{}^4}{R^2 \ell^4} N_P,
\eea
where $N_1$, $N_5$, and $N_P$ count the D1-branes, D5-branes, and units of momentum along the
$S^1$. Reducing on $ T^4 \times  S^1 $, the five-dimensional Newton's constant is
\be
G_5 = \frac{G_{10}}{2 \pi R (2 \pi \ell)^4} =\frac{\pi g^2 \alpha'{}^{4}}{4 R \ell^4},
\ee
and the AdS$_3$ length of the near-horizon region fixed by the D1 and D5 charges \cite{Cvetic:1998xh} is,
\be
l_3 = (Q_1 Q_2)^{\frac{1}{4}} = (N_1 N_5)^{\frac{1}{4}} \sqrt{g} \frac{\alpha'}{\ell} =  (N_1 N_5)^{\frac{1}{4}}  \left( \frac{\sqrt{g} \ell_s}{\ell} \right)  \ell_s.
\ee
For the AdS$_3$  length $l_3$ to be macroscopic, so that the supergravity description is reliable, we require
$N_1 N_5 \gg 1$.

We now take the large $R$ limit
\be
R \gg l_3,
\ee
so that $R$ is the largest scale in the problem. We implement it by scaling $R \to \Lambda R$,
with $\Lambda$ a large dimensionless number, and follow how each quantity behaves as
$\Lambda \to \infty$.

The relations between $Q_1, Q_2$ and $N_1, N_5$ are unaffected, but
holding the momentum $N_P$ fixed suppresses the third charge,
\be
Q_3 \to  \frac{Q_3}{\Lambda^2}.
\ee
The angular momenta require a little more care. The ADM angular momenta are related to the
parameters $J_{3L,3R}$ appearing in the ACS solution by
\be
J_{3L,3R}^\mathrm{ADM} = \frac{\pi}{4 G_5} J_{3L, 3R}.
\ee
The prefactor $\pi/(4 G_5)$ is often set to unity; reinstating it makes the
dependence on $G_5$, and hence on $R$, explicit. Since the $J_{3L,3R}^\mathrm{ADM} $ are quantized and
must be held fixed, while $G_5 \propto 1/R$, the parameters scale as
\be
J_{3L, 3R} \to \frac{J_{3L, 3R}}{\Lambda}.
\ee
Finally, to zoom in on the near-core region we rescale the radial coordinate,
\be
\rho^2 \to \rho^2/\Lambda^2.
\ee

We now follow these scalings through the harmonic functions, working in the
$(\rho, \theta, \phi, \psi)$ coordinates. It is convenient to begin with the base metric parameters $c$,
$d$, and $D$. The first is
\be
c = \frac{J_{3L}}{\sqrt{4 Q_1 Q_2 Q_3}},
\ee
in which the $\Lambda^{-1}$ from $J_{3L}$ is compensated by the $\Lambda^{-1}$ from $\sqrt{Q_3}$,
leaving it invariant,
\be
c \to c .
\ee
The parameter $d$, on the other hand, changes in an essential way,
\be
d \simeq  \frac{J_{3R} \sqrt{Q_3}}{\sqrt{Q_1 Q_2}}  \implies d \to d/\Lambda^2,
\ee
where the explicit factor $\Lambda^{-2}$ comes from the scalings of $J_{3R}$ and $\sqrt{Q_3}$ in
the numerator. Likewise $D = \sqrt{4 Q_1 Q_2 Q_3 - J_{3L}^2}$ scales as
\be
D \to D/\Lambda,
\ee
so that the ratio $J_{3L}/D$ is invariant. Collecting these, the Gibbons-Hawking function grows as
$V \to \Lambda^2 V,$ while the three-dimensional base metric contracts as
$ds^2_{3d} \to \Lambda^{-4} ds^2_{3d}$.
The combination entering the four-dimensional base therefore scales uniformly,
$V ds^2_\mathrm{3d-base} \to \Lambda^{-2} V ds^2_\mathrm{3d-base},$  so that the full four-dimensional flat base simply contracts by $1/\Lambda^2$.

The remaining harmonic functions follow in the same way. Dropping the additive constants where
they are subleading, the $L_I$ behave as
\begin{align}
	L_{1,2} &\sim \mathcal{O}(\Lambda^2), & L_3 &\sim \mathcal{O}(1),&
\end{align}
the constant being retained in $L_3$ precisely because it does not scale. The $K^I$ are
\begin{align}
	K^1 &\sim \frac{Q_2 Q_3}{D} \frac{1}{\rho^2} \sim \mathcal{O}(\Lambda), &
	K^2 &\sim \frac{Q_1 Q_3}{D} \frac{1}{\rho^2} \sim \mathcal{O}(\Lambda), &
	K^3 &\sim \frac{Q_1 Q_2}{D} \frac{1}{\rho^2} \sim \mathcal{O}(\Lambda^3) &
\end{align}
These assemble consistently into the warp functions. The first two are
\begin{align}
	Z_1 &= \frac{K^2 K^3}{V} + L_1 \sim \mathcal{O}(\Lambda^2), &
	Z_2 &= \frac{K^1 K^3}{V} + L_2 \sim \mathcal{O}(\Lambda^2),
\end{align}
each receiving an $\mathcal{O}(\Lambda^2)$ contribution from both terms, so that together with
the $\Lambda^{-2}$ contraction of the base the spatial part of the metric is finite,
\be
\sqrt{Z_1 Z_2} ds^2_\mathrm{4d-base}  \sim  \mathcal{O}(1).
\ee
The function $M$ scales as
\be
M \sim \frac{J_{3L}}{\rho^2} \sim \mathcal{O}(\Lambda)
\ee
while the third warp function stays finite,
\be
Z_3 = \frac{K^1 K^2}{V} + L_3 \sim  \mathcal{O}(1),
\ee

It remains to track the one-forms. The base connection $A$ does not scale, but the rotation one-form $\omega_{BW}$ scales as,
\be
\omega_{BW} \sim \mathcal{O}(\Lambda).
\ee
As a result,  the full one-form $k$ on the four-dimensional base is of order $\Lambda$. This growth is
absorbed by rescaling the time coordinate by the same factor, after which the time fibre is
finite,
\be
- \frac{1}{Z_3 \sqrt{Z_1 Z_2}} (dt + k)^2 \sim  \mathcal{O}(1).
\ee
Finally, the Kaluza-Klein vector along the common circle is
\be
A^3 = - \frac{1}{Z_3} (dt + k) + \frac{K^3}{V} (dz + A) + \xi^3 + dt ,
\ee
which scales as $\mathcal{O}(\Lambda)$; rescaling the circle coordinate accordingly, the
corresponding term in the metric is again finite,
\be
\frac{1}{Z_3 \sqrt{Z_1 Z_2}} (dy + A^3)^2 \sim  \mathcal{O}(1).
\ee

Every block of the ten-dimensional solution thus approaches a finite limit as
$\Lambda \to \infty$, and the rescaled fields define a smooth limiting geometry. In carrying
out these scalings we have used the form of $d$ appropriate to the regime of interest,
\be
d = \frac{J_{3R} \sqrt{Q_3}}{\sqrt{Q_1 Q_2}},
\ee
working throughout in the near-core region $\rho \ll Q_1, Q_2$.

With these new harmonic functions, we can construct the six-dimensional fields following the
procedure of the previous section. The resulting geometry is also obtained in \cite{Larsen:2026sav},
but by a different route: there the $AdS_3\times S^3$ asymptotics are reached either by choosing
four-dimensional multi-center charges and moduli that produce them directly, or by taking a near-horizon
limit of a general non-extremal  six-dimensional black string and  subsequently imposing
the BPS condition on the resulting parameters. Neither route exhibits the large-$R$ limit
connecting the asymptotically flat D1-D5-P index saddle solution to the decoupled throat geometry. The derivation
given here makes this limit explicit, extending to six dimensions the analogous treatment of the 5D black string in \cite{Boruch:2025qdq}.

We  now write the resulting six-dimensional saddle explicitly and analyse its geometry. The
$AdS_3$ factor is spanned by $(t, y, \rho)$ and the three-sphere by $(\theta, \phi, \psi)$; the
coordinate $y$ has period $2\pi R$, and the $AdS_3$ length is $l_3 = (Q_1 Q_2)^{1/4}$. The
metric reads
\be
\begin{aligned} \label{6D_saddle}
	ds^2 = \; &- \frac{J_{3L} J_{3R} \cos(2\theta) + 2 \, l_3^4 \,(\rho^2 - Q_3)}{2l_3^6} dt^2 
	+ \frac{J_{3L} J_{3R} \cos(2\theta) + 2\, l_3^4 \, (Q_3 + \rho^2)}{2l_3^6} dy^2 \\
	&+ \frac{2Q_3}{l_3^2} dt \, dy 
	+ \frac{4l_3^{10} \rho^2}{4l_3^4 \, (J_{3R}^2 \, Q_3 + l_3^4 \, \rho^4) - J_{3L}^2 J_{3R}^2} d\rho^2 \\
	&+ l_3^2 \bigg\{ d\theta^2 + \sin^2\theta \, d\phi^2 + \cos^2\theta \, d\psi^2 \\
	&\hspace{2cm} - \frac{\sin^2\theta}{l_3^4} d\phi \left[ (J_{3L}+J_{3R}) \, dt + (J_{3L}-J_{3R}) \, dy \right] \\
	&\hspace{2cm} + \frac{\cos^2\theta}{l_3^4} d\psi \left[ (J_{3L}-J_{3R}) \, dt + (J_{3L}+J_{3R}) \, dy \right] \bigg\}.
\end{aligned}
\ee

We can readily  bring this into a product of a BTZ black hole and a three-sphere. Recall that the standard rotating BTZ metric,
\be  \label{BTZ_standard}
ds^2_\mathrm{BTZ} = - \frac{(r^2 - r^2_{+})(r^2 - r^2_{-})}{r^2 \, l_3^2} \, d\tau^2 + \frac{r^2 \, l_3^2}{(r^2 - r^2_{+})(r^2 - r^2_{-})} \, dr^2 + r^2 \, \left( d\hat{\phi} + \frac{r_+ \, r_-}{r^2 \, l_3} \, d\tau \right)^2,
\ee
is characterized by its outer and inner horizons, denoted here by $r_+$ and $r_-$, with $\hat{\phi}$
an angle of period $2\pi$. Reaching this form requires a suitable rescaling of the black string
coordinates.

The first step is a shift of the angular coordinates that removes the cross terms,
\bea 
\phi &=& \widetilde \phi  + \kappa_1 \, t + \kappa_2 \, y,  \\
\psi &=& \widetilde \psi - \kappa_2 \, t - \kappa_1 \, y,
\eea
with
\be 
\kappa_1 = \frac{J_{3L} + J_{3R}}{2 l_3^{\, 4}}, \qquad \kappa_2 = \frac{J_{3L} - J_{3R}}{2 l_3^{\, 4}}.
\ee
After this shift the metric simplifies to
\be  \label{6D_no_cross}
\begin{aligned}
	ds^2 = \; &- \frac{J_{3L}^2 + J_{3R}^2 + 4 l_3^{\, 4} \, (\rho^2 - Q_3)}{4 l_3^{\, 6}} dt^2 
	+ \frac{-J_{3L}^2 + J_{3R}^2 + 4 Q_3 \, l_3^{\, 4}}{2 l_3^{\, 6}} dt \, dy \\
	& - \frac{J_{3L}^2 + J_{3R}^2 - 4 l_3^{\, 4} \, (Q_3 + \rho^2)}{4 l_3^{\, 6}} dy^2 
	+ \frac{4 l_3^{\,10} \rho^2}{-J_{3L}^2 J_{3R}^2 + 4 l_3^4 \left(J_{3R}^2 \, Q_3 + \rho^4 \, l_3^{\, 4} \right)} d\rho^2 \\
	& + l_3^2 \, \left( d\theta^2  + \cos^2{\theta} \, d\widetilde{\psi}^{\, 2} + \sin^2{\theta} \, d\widetilde{\phi}^{\,2} \right),
\end{aligned}
\ee
which is manifestly a product.
 The last line is a three-sphere of radius $l_3$, while the
$(t, y, \rho)$ part is the rotating BTZ geometry, its rotation residing in the $t$–$y$ cross
term.  In the Euclidean frame the time coordinate is periodic and scales as $R$.
We rescale
\be \label{ty_scalings} 
y = \hat{\phi} \, R, \qquad t = \frac{R}{l_3} \, \tau.
\ee
	The physical meaning of these rescalings is as follows. In the large $R$ limit,
	we write the throat coordinates, stripped of the $\Lambda$ scaling, as
	\be \label{eq:hatted_coords}
	t = \Lambda \, \hat t \,, \qquad y = \Lambda \, \hat y \,.
	\ee
	The asymptotic period of the $y$-circle after the scaling $R \to \Lambda R$ is
	$2\pi (\Lambda R)$, which becomes $2 \pi R$ in the $\hat y$ coordinate. In the same
	way, in the Euclidean section, if the time circle of the asymptotically flat index
	saddle has period $\beta$, then the throat time coordinate has period
	$\hat \beta = \beta/\Lambda$.
	Holding $\hat\beta$ fixed as $\Lambda \to \infty$ is therefore the statement that
	\be \label{eq:finiteTdecoupling}
	\beta \to \infty \,, \qquad R \to \infty \,, \qquad \frac{\beta}{R} \quad \text{fixed} \,,
	\ee
	that is, both cycles of the boundary torus are taken to infinity at the same
	rate. This is precisely the finite-temperature decoupling limit used in
	\cite{Boruch:2025qdq} for the five-dimensional black string. It remains to identify the inner and outer horizons, which we read off from the $g_{\rho\rho}$
component. We find
\bea 
r_+ &=& (D - i J_{3R}) \, \frac{R}{2 l_3^3}, \label{BTZ_o_horizon}\\
r_- &=& (D + i J_{3R}) \, \frac{R}{2 l_3^3}. \label{BTZ_i_horizon}
\eea
with $D = \sqrt{4 Q_1 Q_2 Q_3 - J_{3L}^2}$ as above.
Recall the restrictions on the parameters: we take $4 Q_1 Q_2 Q_3 > J_{3L}^2$, so that
\be
D = \sqrt{4 Q_1 Q_2 Q_3 - J_{3L}^2} > 0,
\ee 
and $J_{3R}$ lies on the
positive imaginary axis, $-i J_{3R} > 0$.  Re-expressing the $AdS_3$ part in terms of
these horizons and the rescaled variables \eqref{ty_scalings} brings it precisely to the
standard BTZ form \eqref{BTZ_standard}. Equivalently, written in terms of the BTZ mass and
angular momentum,
\be 
ds^2_\mathrm{BTZ} = -\left( \frac{r^2}{l_3^2} - M_3 + \frac{J_3^2}{4 r^2} \right) d\tau^2 + \left( \frac{r^2}{l_3^2} - M_3 + \frac{J_3^2}{4 r^2} \right)^{-1} dr^2 + r^2 \left( d \hat \phi + \frac{J_3}{2 r^2} d\tau \right)^2.
\ee
we identify
\bea
M_3 & =& (D^2 - J_{3R}^2) \, \frac{R^{\, 2}}{2 l_3^{\, 8}}, \\
J_3 & =& (D^2 + J_{3R}^2) \, \frac{R^{\, 2}}{2 l_3^{\, 7}}.
\eea

It is illuminating to phrase this thermodynamic data in the language of the dual
CFT$_2$. Under the AdS$_3$/CFT$_2$ dictionary the combinations $l_3 M_3 \pm J_3$ are proportional
to the left- and right-moving conformal weights $L_0$ and $\bar{L}_0$. The left-moving weight
$l_3 M_3 + J_3 \propto D^2 = 4 Q_1 Q_2 Q_3 - J_{3L}^2$ is independent of $J_{3R}$: it is fixed by the
charges and the left R-charge $J_{3L}$, and controls the Cardy degeneracy
$\sqrt{L_0} \propto D$ reproduced by the index. The right-moving weight, on the other
hand, is governed by $J_{3R}$ alone, $l_3 M_3 - J_3 \propto -J_{3R}^2$. Since $J_{3R}$ is purely
imaginary on the index saddle, $J_{3R}^2 < 0$ and hence $l_3 M_3 > J_3$, so the right-moving weight
is strictly positive: the right sector is thermally excited as well, and the saddle carries a
finite temperature in both chiral sectors rather than freezing the right-movers into their
Ramond ground state, as the extremal BPS black hole does. That supersymmetry nonetheless
survives was clarified recently by Larsen and Sharma~\cite{Larsen:2026sav}, building upon the observations in \cite{Larsen:2021wnu, Larsen:2025jqo}. 

Since the index saddle is a solution of the Euclidean equations of motion, we close this
section by recording the Wick rotation $\tau = -i \, t_E$, the spatial circle $y$  being left untouched. Under this continuation the BTZ factor \eqref{BTZ_standard} becomes the
standard Euclidean rotating BTZ metric
\be \label{eq:BTZ_E}
ds^2_{\mathrm{BTZ,E}} = \frac{(r^2 - r^2_{+})(r^2 - r^2_{-})}{r^2 \, l_3^2} \, dt_E^2
+ \frac{r^2 \, l_3^2}{(r^2 - r^2_{+})(r^2 - r^2_{-})} \, dr^2
+ r^2 \left( d\hat\phi - i \, \frac{r_+ r_-}{r^2 \, l_3} \, dt_E \right)^2 ,
\ee
with \be
r_+ = (D + J) \, \frac{R}{2 l_3^3} \,, \qquad r_- = (D - J) \, \frac{R}{2 l_3^3} \,,
\qquad J \equiv - i J_{3R} > 0 \,.
\ee
We note that both radii $r_\pm$ are real, whereas a real Euclidean rotating BTZ solution requires $r_-$ to be purely
imaginary. The Euclidean section is therefore complex.

\section{Conclusions}
\label{sec:conclusions}

In this paper, we constructed the index saddle for the D1-D5-P black string and used it to obtain, via an explicit decoupling limit, the BTZ$~\times~S^3$ saddle that computes the index of the D1-D5 CFT. It is worth revisiting the steps of this construction (Sections~\ref{sec:4charge}--\ref{sec:decoupling}). The four-charge index saddle of Section~\ref{sec:4charge} is an instance of the new attractor mechanism~\cite{Boruch:2023gfn}, independent of the BMPV black hole. The uplift chain of Section~\ref{sec:D1D5saddle} shows that the four-dimensional and the five- and six-dimensional index-saddles are essentially the same gravitational solution viewed in different duality frames. The decoupling limit of Section~\ref{sec:decoupling} then delivers an explicit gravitational saddle reproducing the index of the D1-D5 CFT.

Our construction has two gaps that we wish to record here explicitly. First, the matching between the four-dimensional index saddle and the ACS saddle in Section~\ref{sec:4d5dconnection} was carried out only at $J_{3L}=0$. The four-charge non-extremal solution of Section~\ref{sec:4charge} does not carry the electric charge of the Kaluza-Klein vector $A^0$ which becomes $J_{3L}$ under the 4D-5D connection. As noted, turning it on is a technical complication, but we see no conceptual barrier to completing the construction at $J_{3L} \neq 0$. 

Second, the $\beta$-independence of the six-dimensional on-shell action, argued in Section~\ref{sec:indexsaddleD1D5}, rests on dimensionally reducing the D1-D5-P solution back down to five dimensions and identifying the result with the  on-shell action of the ACS saddle. This dimensional reduction requires an electromagnetic duality on one of the field strengths, whose effect on the action we have not tracked. We expect this duality to act trivially on the on-shell action, consistent with the new attractor mechanism, but we have not verified this explicitly.

While this work was in progress, Larsen and Sharma~\cite{Larsen:2026sav} constructed closely related $AdS_3 \times S^3 \times T^4$ saddles by two routes: (i) starting from four-dimensional multi-center data and uplifting to six dimensions, and (ii)  by taking the near-horizon limit of a non-extremal six-dimensional black string before imposing the BPS condition. The route taken in this paper is different. We obtained it as the large-radius decoupling limit of an explicit, asymptotically flat, D1-D5-P index saddle.

More broadly, this work is part of a growing effort to understand, directly in gravity, why the logarithm of the supersymmetric index of a string-theory black hole agrees with its macroscopic entropy. The new attractor mechanism, the uplift chain connecting four, five, and six dimensions, and the decoupling limits relating asymptotically flat saddles to their near-horizon AdS duals together give a reasonably complete gravitational picture for the BMPV and D1-D5-P systems studied here. Many questions remain. Perhaps the most pressing is to understand the physical relevance of the kind of solutions constructed here. Our saddles are generically complex, and there is at present no settled criterion for which complex saddles should be admitted into the gravitational path integral. It will be interesting to explore our solutions in relation to the different proposals that address this question: (i) the Kontsevich-Segal-Witten allowability criterion~\cite{Kontsevich:2021dmb,Witten:2021nzp}, (ii) the Lorentzian gravitational path integral approach of Kolanowski and Marolf~\cite{Kolanowski:2026gii}, and (iii) the minisuperspace approach of Mahajan and Singhi~\cite{Mahajan:2025bzo, Singhi:2025rfy, Ailiga:2025osa}, which asks whether the defining contour of the simplified path integral can be deformed into the steepest descent of the complex saddles. We hope to return to these questions in the future.

\bigskip

\noindent{\bf Acknowledgements:}  We thank Iosif Bena, Jan Boruch, Davide Cassani, Roberto Emparan, Silvia Georgescu, Finn Larsen, Sameer Murthy, and especially David Turton for discussions at the GGI Workshop ``Pathways to Quantum Black Holes: from Effective Theories to Exact Methods.'' During the workshop, we became aware that several groups were working on related problems, and we thank them for generously sharing their ideas and details of their constructions. The work of A.V.\ was partly supported by SERB Core Research Grant CRG/2023/000545. P.S.\ acknowledges the support of the Department of Atomic Energy, Government of India, under project no. RTI4019. P.S.\ thanks Chennai Mathematical Institute (CMI) for the warm hospitality during the final stages of this work. K.K.N.\ thanks the organizers of the ``Spring School on Superstring Theory and Related Topics'' held at ICTP for giving the opportunity to present part of this work as a poster. K.K.N.\ also thanks the strings group at IIT Indore for their hospitality during the completion of this work. We acknowledge the support from the Infosys Foundation to CMI.

\appendix

\section{Supersymmetric limit of the non-extremal D1-D5-P black string}
\label{5d6dbpsapp}
Jejjala, Madden, Ross and Titchener (JMaRT)~\cite{Jejjala:2005yu} wrote down the non-extremal D1-D5-P black string. Reducing this six-dimensional solution on the $y$-circle returns us directly to the non-extremal five-dimensional metric that is the starting point of the Anupam-Chowdhury-Sen (ACS) saddle construction~\cite{Anupam:2023yns}. This is useful because it lets us instead take the ACS supersymmetric limit directly on the six-dimensional solution, and obtain the six-dimensional finite temperature supersymmetric black string. This is precisely the index saddle constructed in the main text of the paper. This appendix thus provides another consistency check on the full logic of the paper.

\subsection{Non-extremal D1-D5-P black string}

JMaRT write the non-extremal D1-D5-P black string metric as\footnote{Compared to \cite{Jejjala:2005yu} we have flipped the sign of $y$ for later convenience.}
\begin{align}
	ds^2 &= -\frac{f}{\sqrt{\tilde H_1\tilde H_5}}(dt^2-dy^2)+\frac{M}{\sqrt{\tilde H_1\tilde H_5}}(-s_p\,dy-c_p\,dt)^2 \nonumber\\
	&+\sqrt{\tilde H_1\tilde H_5}\left(\frac{r^2dr^2}{(r^2+a_1^2)(r^2+a_2^2)-Mr^2}+d\theta^2\right) \nonumber\\
	&+\left(\sqrt{\tilde H_1\tilde H_5}-(a_2^2-a_1^2)\frac{(\tilde H_1+\tilde H_5-f)\cos^2\theta}{\sqrt{\tilde H_1\tilde H_5}}\right)\cos^2\theta\,d\psi^2 \nonumber\\
	&+\left(\sqrt{\tilde H_1\tilde H_5}+(a_2^2-a_1^2)\frac{(\tilde H_1+\tilde H_5-f)\sin^2\theta}{\sqrt{\tilde H_1\tilde H_5}}\right)\sin^2\theta\,d\phi^2 \nonumber\\
	&+\frac{M}{\sqrt{\tilde H_1\tilde H_5}}(a_1\cos^2\theta\,d\psi+a_2\sin^2\theta\,d\phi)^2 \nonumber\\
	&+\frac{2M\cos^2\theta}{\sqrt{\tilde H_1\tilde H_5}}\big[(a_1c_1c_5c_p-a_2s_1s_5s_p)dt-(a_2s_1s_5c_p-a_1c_1c_5s_p)dy\big]d\psi \nonumber\\
	&+\frac{2M\sin^2\theta}{\sqrt{\tilde H_1\tilde H_5}}\big[(a_2c_1c_5c_p-a_1s_1s_5s_p)dt-(a_1s_1s_5c_p-a_2c_1c_5s_p)dy\big]d\phi, 
	\label{eq:JMaRTmetric}
\end{align}
where $c_i\equiv\cosh\delta_i,\ s_i\equiv\sinh\delta_i$ and
\begin{equation}
	\tilde H_i = f+M\sinh^2\delta_i\,, \qquad f = r^2+a_1^2\sin^2\theta+a_2^2\cos^2\theta\,, \qquad i=1,5,p\,.
	\label{eq:Htilde}
\end{equation}
The dilaton and the Ramond-Ramond two-form supporting this configuration are~\cite{Jejjala:2005yu}
\begin{equation}
	e^{2\Phi} = \frac{\tilde H_1}{\tilde H_5}\,,
	\label{eq:dilaton}
\end{equation}
\begin{align}
	C_2 &= \frac{M\cos^2\theta}{\tilde H_1}\big[(a_2c_1s_5c_p-a_1s_1c_5s_p)dt - (a_1s_1c_5c_p-a_2c_1s_5s_p)dy\big]\wedge d\psi \nonumber\\
	&+\frac{M\sin^2\theta}{\tilde H_1}\big[(a_1c_1s_5c_p-a_2s_1c_5s_p)dt - (a_2s_1c_5c_p-a_1c_1s_5s_p)dy\big]\wedge d\phi \nonumber\\
	&+\frac{Ms_1c_1}{\tilde H_1}dt\wedge dy-\frac{Ms_5c_5}{\tilde H_1}(r^2+a_2^2+Ms_1^2)\cos^2\theta\,d\psi\wedge d\phi\,. \label{eq:C2}
\end{align}

The five free parameters $(M,a_1,a_2,\delta_1,\delta_5,\delta_p)$ are exactly those of the ACS presentation, with the dictionary
\begin{equation}
	M=2m\,, \qquad a_1=-l_2\,, \qquad a_2=-l_1\,, \qquad \delta_1=\delta_{e_1}\,, \qquad \delta_5=\delta_{e_2}\,, \qquad \delta_p=\delta_{e_3}\,.
	\label{eq:dictionary}
\end{equation}

We now reduce \eqref{eq:JMaRTmetric} on the $y$-circle. Writing the reduction in the standard Kaluza-Klein form
\begin{equation}
	ds^2_\mathrm{JMaRT} = F^{-1/3}ds^2_{5}+F(dy+A^3)^2\,,
	\label{eq:KKform}
\end{equation}
the coefficient of $dy^2$ in \eqref{eq:JMaRTmetric} fixes $F$ directly,
\begin{equation}
	F = \frac{f+Q_p}{\sqrt{\tilde H_1\tilde H_5}}=\frac{\tilde H_p}{\sqrt{\tilde H_1\tilde H_5}}\,,
	\label{eq:Fdef}
\end{equation}
and the $dt\,dy$, $d\phi\,dy$, $d\psi\,dy$ cross-terms fix the Kaluza-Klein gauge field $A^3$,
\begin{align}
	A^3 &= \frac{M}{\tilde H_p}k_p\,, \\
	k_p &= s_pc_1c_5\left(\frac{c_p}{c_1c_5}dt+a_2\sin^2\theta\,d\phi+a_1\cos^2\theta\,d\psi\right)-c_ps_1s_5\left(a_1\sin^2\theta\,d\phi+a_2\cos^2\theta\,d\psi\right).
	\label{eq:Apdef}
\end{align}
With $F$ and $A^3$ so determined, the remaining five-dimensional metric is, by construction, free of $y$:
\begin{equation}
	ds_5^2 = F^{1/3}\Big[ds^2_\mathrm{JMaRT}-F(dy+A^3)^2\Big]\,.
	\label{eq:ds5def}
\end{equation}
Substituting \eqref{eq:JMaRTmetric}--\eqref{eq:Htilde} into \eqref{eq:ds5def} and using the dictionary~\eqref{eq:dictionary}, $ds_5^2$ is exactly the non-extremal ACS metric, eq.~(A.1) of \cite{Anupam:2023yns}.

\subsection{Finite temperature supersymmetric limit}

Following ACS, we now take $m \to 0$ holding the
charges $Q_i$ and the angular momenta $J_{3L}, J_{3R}$ fixed. This limit yields
the index saddle, as can be seen as follows. For the non-extremal solution of the
previous subsection, the angular velocity conjugate to $J_{3R}$ follows from the
entropy $S_0$ as $\beta \Omega_R = \partial S_0 / \partial J_{3R}$. As shown by
ACS~\cite{Anupam:2023yns} (their eqs.~(A.7)--(A.9)), the limit above is precisely
the locus
\be
\beta \, \Omega_R = -\, 2 \pi i \,.
\ee
The Euclidean solution
constructed below therefore contributes to the supersymmetric index, rather than to the thermal partition function. Under the dictionary~\eqref{eq:dictionary}, this is the limit $\delta_1,\delta_5,\delta_p\to\infty$, $M\to0$, with
\begin{equation}
	Q_i = M\sinh\delta_i\cosh\delta_i\,, \qquad i=1,5,p
\end{equation}
held fixed, so that
\begin{align}
	\cosh^2\delta_i &= \frac12+\sqrt{\frac14+\frac{Q_i^2}{M^2}}\,, & \sinh^2\delta_i &= -\frac12+\sqrt{\frac14+\frac{Q_i^2}{M^2}}\,.
\end{align}
Translating ACS's eqs.~(A.11)--(A.12) through the dictionary~\eqref{eq:dictionary} gives the corresponding limiting form of $a_1,a_2$,
\begin{align}
	a_1+a_2 &= -\frac{2J_{3R}\sqrt{Q_1Q_5Q_p}}{\sqrt M\,Q_S}\,,  &
	a_1-a_2 &= \frac{J_{3L}\sqrt M}{2\sqrt{Q_1Q_5Q_p}}\,, 
	\label{eq:apm}
\end{align}
where 
\be
Q_S\equiv Q_1Q_5+Q_5Q_p+Q_pQ_1\,.
\ee
This quantity is the same as the $Q_S$ defined in~\eqref{eq:Q_S} with dictionary $Q_1,Q_2,Q_3\to Q_5,Q_1,Q_p$. Next, exactly as in ACS's eq.~(A.17), we define the finite radial variable
\begin{equation}
	\rho^2 \equiv r^2-\tfrac12(M-a_1^2-a_2^2)\,,
\end{equation}
which stays finite as $a_1,a_2$ diverge.

Several of the metric coefficients below are built from $a_1,a_2$ multiplying $c_i,s_i$, and each such combination is individually divergent as $M\to0$; only after combining \eqref{eq:apm} with the subleading piece of $\cosh\delta_i,\sinh\delta_i$ does a finite limit emerge, so $\delta_i\to\infty$ cannot simply be substituted termwise. With this understood, the limit gives
\begin{equation}
	f \to \rho^2+\frac{J_{3L}J_{3R}}{2Q_S}\cos(2\theta)\,, \qquad \tilde H_i \to f+Q_i\,,
\end{equation}
\begin{equation}
	(r^2+a_1^2)(r^2+a_2^2)-Mr^2 \to \rho^4+\frac{J_{3R}^2(4Q_1Q_5Q_p-J_{3L}^2)}{4Q_S^2}\,.
\end{equation}

The components of the six-dimensional metric, become in this limit:
\begin{align}
	g_{tt} &= -\frac{f-Q_p}{\sqrt{\tilde H_1\tilde H_5}}\,, &
	g_{ty} &= \frac{Q_p}{\sqrt{\tilde H_1\tilde H_5}}\,, \\
	g_{yy} &= \frac{f+Q_p}{\sqrt{\tilde H_1\tilde H_5}}\,, &
	g_{\theta\theta} &= \sqrt{\tilde H_1\tilde H_5}\,, 
\end{align}
\begin{align}
	g_{\rho\rho} &= \frac{\sqrt{\tilde H_1\tilde H_5}\,\rho^2}{\rho^4+\dfrac{J_{3R}^2(4Q_1Q_5Q_p-J_{3L}^2)}{4Q_S^2}}\,,  \\
	g_{\psi\psi} &= \frac{\cos^2\theta}{\sqrt{\tilde H_1\tilde H_5}}\left[\tilde H_1\tilde H_5+\cos^2\theta\left(\frac{J_{3R}^2Q_1Q_5Q_p}{Q_S^2}-\frac{J_{3L}J_{3R}(f+Q_1+Q_5)}{Q_S}\right)\right]\,, \\
	g_{\phi\phi} &= \frac{\sin^2\theta}{\sqrt{\tilde H_1\tilde H_5}}\left[\tilde H_1\tilde H_5+\sin^2\theta\left(\frac{J_{3R}^2Q_1Q_5Q_p}{Q_S^2}+\frac{J_{3L}J_{3R}(f+Q_1+Q_5)}{Q_S}\right)\right]\,, \\
	g_{\phi\psi} &= \frac{J_{3R}^2Q_1Q_5Q_p}{Q_S^2}\,\frac{\cos^2\theta\sin^2\theta}{\sqrt{\tilde H_1\tilde H_5}}\,, \\
	g_{t\psi} &= \frac{J_{3L}-J_{3R}}{2}\,\frac{\cos^2\theta}{\sqrt{\tilde H_1\tilde H_5}}\,, \\
	g_{y\psi} &= \frac12\left(J_{3L}-J_{3R}\frac{Q_1Q_p+Q_5Q_p-Q_1Q_5}{Q_S}\right)\frac{\cos^2\theta}{\sqrt{\tilde H_1\tilde H_5}}\,, \\
	g_{t\phi} &= -\frac{J_{3L}+J_{3R}}{2}\,\frac{\sin^2\theta}{\sqrt{\tilde H_1\tilde H_5}}\,, \\
	g_{y\phi} &= -\frac12\left(J_{3L}+J_{3R}\frac{Q_1Q_p+Q_5Q_p-Q_1Q_5}{Q_S}\right)\frac{\sin^2\theta}{\sqrt{\tilde H_1\tilde H_5}}\,.
\end{align}
This is the metric of the finite-temperature, six-dimensional supersymmetric black string.

We now take the same limit of the Ramond-Ramond two-form~\eqref{eq:C2}. The $dt\wedge dy$ and $d\psi\wedge d\phi$ terms only need the exact charge relation $Ms_1c_1=Q_1$, $Ms_5c_5=Q_5$, together with
\begin{equation}
	r^2+a_2^2+Ms_1^2 \to \rho^2+Q_1+\frac{J_{3L}J_{3R}}{2Q_S}\,.
\end{equation}
The remaining four terms are, as for the metric, built from $a_1,a_2$ multiplying $c_i,s_i$, and require the same care with the subleading pieces of $\cosh\delta_i,\sinh\delta_i$ described above. The components of the two-form in this limit become:
\begin{align}
	C_{ty} &= \frac{Q_1}{\tilde H_1}\,, \\
	C_{t\psi} &= \frac{\cos^2\theta}{2\tilde H_1}\left(-J_{3L}-J_{3R}\frac{Q_1Q_5+Q_5Q_p-Q_1Q_p}{Q_S}\right)\,, \\
	C_{y\psi} &= \frac{\cos^2\theta}{2\tilde H_1}\left(-J_{3L}+J_{3R}\frac{Q_1Q_5+Q_1Q_p-Q_5Q_p}{Q_S}\right)\,, \\
	C_{t\phi} &= \frac{\sin^2\theta}{2\tilde H_1}\left(J_{3L}-J_{3R}\frac{Q_1Q_5+Q_5Q_p-Q_1Q_p}{Q_S}\right)\,, \\
	C_{y\phi} &= \frac{\sin^2\theta}{2\tilde H_1}\left(J_{3L}+J_{3R}\frac{Q_1Q_5+Q_1Q_p-Q_5Q_p}{Q_S}\right)\,, \\
	C_{\psi\phi} &= -\frac{Q_5}{\tilde H_1}\left(\rho^2+Q_1+\frac{J_{3L}J_{3R}}{2Q_S}\right)\cos^2\theta\,.
\end{align}

The dilaton~\eqref{eq:dilaton} has no $y$-dependence and involves no singular combination of $a_1,a_2$; it is simply
\begin{equation}
	e^{2\Phi} = \frac{f+Q_1}{f+Q_5}\,.
\end{equation}

\section{Emparan-Maccarrone zooming in procedure}
\label{4d5dapp}

Emparan and Maccarrone (EM)~\cite{Emparan:2007en} showed that a general non-extremal rotating dyonic
Kaluza-Klein black hole, in a limit where its magnetic charge is sent to infinity, reduces to
the general vacuum Myers-Perry black hole in five dimensions. The Kaluza-Klein black hole
becomes a Myers-Perry black hole sitting at the tip of the Taub-NUT space.
In this appendix, we extend their zooming-in procedure to the charged setting relevant for
the four-dimensional black hole of Section~\ref{sec:4charge}. The starting point is exactly of the type considered in~\cite{Emparan:2007en}; except that \textit{our black hole does not carry the Kaluza-Klein electric charge}. The black hole of Section~\ref{sec:4charge} carries the same Kaluza-Klein magnetic charge, parametrized by
$\gamma_4$, that is sent to infinity in the zoom-in limit. In addition, it carries three other 
electric charges parametrized by $\gamma_1,\gamma_2,\gamma_3$.  Taking the limit $\gamma_4\to\infty$ on this solution, with
$\gamma_1,\gamma_2,\gamma_3$ held fixed, decompactifies the Kaluza-Klein circle and yields
the general non-extremal five-dimensional black hole with \textit{special values of the angular momenta}. In this appendix,  we carry out this limit.

\subsection{$\gamma_1=\gamma_2=\gamma_3=0$}
\label{sec:EM_check}

Before applying the zoom-in procedure to the full four-charge solution,
it is worth carrying it out in detail on the simplest possible truncation, since the general
case is a technical extension of it. Working through this case first lets us spell out the
key points. We start from the scalars of eq.~\eqref{sca1}--\eqref{sca3}. From these we determine the
function controlling the radius of the Kaluza-Klein circle,
\begin{equation}
	\tilde{f} = \exp\Big(-\frac{\varphi_1+\varphi_2+\varphi_3}{3}\Big) = \frac{W}{F_1 F_2 F_3}, \qquad F_i = (r^2 + a^2 \cos^2 \theta + g_i)^{\frac{1}{3}}. \label{tildef}
\end{equation}
Given the 4D metric \eqref{CC_non-ext}, the five-dimensional metric is obtained through
the standard Kaluza-Klein uplift formula
\begin{equation}
	ds_5^2 = \tilde{f}^2 (dz-A^0)^2 +\tilde{f}^{-1} ds_4^2, \qquad   A^0 = \zeta^4(dt+\omega_3)+A^4_{3D},
	\label{eq:upliftcheck}
\end{equation}
with $z$ the Kaluza-Klein circle (not to be confused with the sixth dimension $y$).

Setting $\gamma_1=\gamma_2=\gamma_3=0$ removes all electric charge and leaves a three-parameter
$(m,a,\gamma_4)$ family carrying only Kaluza--Klein magnetic charge. Defining
\begin{equation}
	P\equiv r^2+a^2\cos^2\theta\,,\qquad F\equiv P+2mr\,s_4^2 \;\big(=(F_1)^3=(F_2)^3=(F_3)^3\big)\,,
	\label{eq:PFdef}
\end{equation}
one finds $g_1=g_2=g_3=2mr\,s_4^2$ and, from \eqref{tildef}, $W^2= PF$, so that
\begin{equation}
	\tilde f=\frac{W}{F_1F_2F_3}=\sqrt{\frac{P}{F}}\,.
	\label{eq:ftildecheck}
\end{equation}
The four-dimensional metric \eqref{CC_non-ext} reduces to
\begin{equation}
	ds_4^2=-\frac{\Delta_\theta}{\sqrt{PF}}\big(dt+\omega_3\big)^2
	+\sqrt{PF}\left[\frac{dr^2}{\Delta}+d\theta^2+\frac{\Delta}{\Delta_\theta}\sin^2\theta\,d\phi^2\right]\,,
	\label{eq:ds4check}
\end{equation}
with
\begin{equation}
	\Delta\equiv r^2-2mr+a^2\,,\qquad \Delta_\theta\equiv r^2-2mr+a^2\cos^2\theta\,,\qquad
	\omega_3=\frac{2ma\,c_4\,r\sin^2\theta}{\Delta_\theta}\,d\phi\,.
	\label{eq:Deltacheck}
\end{equation}
The gauge fields $A_i$ vanish identically for $i=1,2,3$, while the Kaluza--Klein vector
$A^0=\zeta^4(dt+\omega_3)+A^4_{3D}$ becomes
\begin{equation}
	\zeta^4=-\frac{2am\,s_4\cos\theta}{P}\,,\qquad
	A^4_{3D}=\frac{2m\,c_4 s_4\,\Delta\,\cos\theta}{\Delta_\theta}\,d\phi\,.
	\label{eq:zeta4check}
\end{equation}

The metric \eqref{eq:ds4check}--\eqref{eq:zeta4check} is the
Rasheed--Larsen rotating Kaluza--Klein black hole with vanishing electric charge. Concretely,
identifying $\alpha_{EM}=a$, $m_{EM}=m$, $r_{EM}=r$ and $p_{EM}=2mc_4^2$, one finds
\begin{equation}
	P=H_q\,,\qquad F=H_p\,,\qquad \Delta=\Delta_{EM}\,,\qquad \Delta_\theta=\Delta_{\theta,EM}\,,
\end{equation}
where $H_p,H_q,\Delta_{EM},\Delta_{\theta,EM}$ are the functions of Emparan--Maccarrone's
eq.~(A.1) specialised to $q=2m$. Uplifting using \eqref{eq:ftildecheck} gives EM eq.~(A.1) at $q=2m$.

Because $q=2m$ kills the Kaluza--Klein electric charge, consistency of
EM eqs.~(A.14) and (A.16) forces their two rotation parameters to satisfy
\be
b_{EM}=-a_{EM}.
\ee Only a single rotation parameter survives the zoom-in. The Kaluza-Klein circle itself must be rescaled as
it decompactifies,
\be
z = p\,\psi\,,
\label{eq:zpsi}
\ee
Trading $\gamma_4$ for
$p\equiv 2mc_4^2\to\infty$ and sending $m,r,a\to0$ holding fixed
\begin{equation}
	pm=\frac{\mu}{8}\,,\qquad pr=\frac14(\rho^2+b^2)\,,\qquad pa=-\frac14\sqrt\mu\,b\,,
	\label{eq:zoomdictcheck}
\end{equation}
the identification $a_{EM}=-b$ and the angle map $\theta=2\eta$, $\psi=\xi-\zeta$,
$\phi=\xi+\zeta$ turn the metric above into
\bea
ds_5^2&=&-dt^2+\frac{\mu}{\rho^2+b^2}\Big(dt+b\sin^2\eta\,d\xi+b\cos^2\eta\,d\zeta\Big)^2 \nonumber\\
&&+(\rho^2+b^2)\!\left(\frac{d\rho^2}{\tilde\Delta}+d\eta^2\right) \nonumber\\
&&+(\rho^2+b^2)\sin^2\eta\,d\xi^2+(\rho^2+b^2)\cos^2\eta\,d\zeta^2
\label{eq:finalMPcheck}
\eea
with
\begin{equation}
	\tilde\Delta\equiv\frac{(\rho^2+b^2)^2-\mu\rho^2}{\rho^2}\,.
\end{equation}
Equation \eqref{eq:finalMPcheck} is the equal angular momentum five-dimensional Myers--Perry
metric.

\subsection{General $\gamma_1,\gamma_2,\gamma_3$}
\label{sec:EM_general}

We now repeat the zoom-in of \S\ref{sec:EM_check}, keeping the three electric
parameters $\gamma_1,\gamma_2,\gamma_3$ generic. The aim is correspondingly richer. We wish to recover the general non-extremal,
three-charge, Cveti\v{c}-Youm black hole~\cite{Cvetic:1996xz, Cvetic:1996kv}---as written in \cite[eq.~(A.1)--(A.2)]{Anupam:2023yns})---restricted to equal
rotation parameters $l_1=l_2$. 

The zoom-in dictionary is exactly~\eqref{eq:zoomdictcheck}.  Together with the angle map already used in \S\ref{sec:EM_check},
$\theta=2\eta,\ \psi=\xi-\zeta,\ \phi=\xi+\zeta$. Matching every component of
the zoomed metric against~\cite{Anupam:2023yns} fixes the complete
dictionary,
\begin{align}
	\rho &= r_{ACS}\,, & \mu &= 2m\,,&  b& =-l_1=-l_2\,, \\
	\eta &= \theta_{ACS}\,,& \xi &=\phi_{ACS}\,,
	& \zeta& =\psi_{ACS}\,,
	\label{eq:fulldict}
\end{align}
togehter with $\gamma_i =\delta_{e_i}\ (i=1,2,3)$.
On the right-hand side $r_{ACS},m,l_1,l_2,\theta_{ACS},\phi_{ACS},\psi_{ACS}$ are
the parameters and coordinates of~\cite{Anupam:2023yns}. The relative sign
$b=-l_1$ costs nothing in the components built only from $l_1^2,l_2^2,l_1l_2$
($g_{rr},g_{\theta\theta},g_{\phi\phi},g_{\psi\psi},g_{\phi\psi}$), but is
essential to obtain the correct sign of the terms
$g_{t\phi},g_{t\psi}$, which are linear in the rotation parameters.

\paragraph{Example: the radial component.} As an illustration, consider
$g_{rr}$. The uplift formula~\eqref{eq:upliftcheck} gives
\be
g_{rr}^{(5d)} = \tilde f^{-1}\cdot\frac{W}{r^2-2mr+a^2}
= \frac{F_1F_2F_3}{r^2-2mr+a^2}\,,
\ee
the factor of $W$ cancelling between $\tilde f^{-1}=F_1F_2F_3/W$ and the
four-dimensional metric. Substituting the zoom-in dictionary into the general
scalars $g_i$, cf.~eq.~\eqref{sca2}, the terms proportional to $rs_i^2$ and
$s_i^2s_4^2$ vanish as $p\to\infty$, while
\be
g_i \;\longrightarrow\; \tfrac14\big[(\rho^2+b^2)+\mu\sinh^2\gamma_i\big]
\equiv \tfrac14\tilde H_i\,,
\ee
so that $F_1F_2F_3\to\frac14(\tilde H_1\tilde H_2\tilde H_3)^{1/3}$. Together
with $r^2-2mr+a^2\to[(\rho^2+b^2)^2-\mu\rho^2]/16p^2$ and 
$dr=(\rho/2p)\,d\rho$, this gives
\be
g_{\rho\rho}\,d\rho^2 =
\frac{\rho^2\big(\tilde H_1\tilde H_2\tilde H_3\big)^{1/3}}
{(\rho^2+b^2)^2-\mu\rho^2}\,d\rho^2\,,
\qquad \tilde H_i \equiv \rho^2+b^2+\mu\sinh^2\gamma_i\,.
\label{eq:grrgeneral}
\ee
This is precisely the $g_{rr}$ term of~\cite{Anupam:2023yns}. 

\paragraph{The remaining components.} The same steps, that is, substitute the
dictionary into the relevant piece of the general non-extremal solution of
\S\ref{sec:4charge}, take $p\to\infty$, and compare
against~\cite{Anupam:2023yns}, go through for every other component. The $\phi_{ACS}$ and $\psi_{ACS}$ genuinely mix. The zoomed solution
is naturally diagonal in $\phi$ and the rescaled Kaluza-Klein angle $\psi = z/p$,
while metric in~\cite{Anupam:2023yns} is naturally diagonal in $\phi_{ACS}$, $\psi_{ACS}$. Matching the angular sector and the temporal sector therefore requires combining
the zoomed components through this recombination; once this is done, every
one of them agrees with~\cite{Anupam:2023yns} exactly. We have verified all eight independent metric components in this way.
The vector and scalar fields can similarly be computed and shown to match the expressions given in \cite{Adhikari:2024zif}. They are,
\begin{align}
	h_i = \frac{(\tilde{H_1} \tilde{H}_2\tilde{H}_3)^{\frac{1}{3}}}{\tilde{H}_i},
\end{align}
and
\begin{align}
\tilde{A}_i = \frac{\mu}{\tilde{H}_1} \left(c_i s_i dt + b \left(\frac{s_1 c_{123}}{c_1}- \frac{c_1 s_{123}}{s_1}\right) \left(\cos^2 \eta \, d\zeta + \sin^2 \eta \, d\xi\right)\right).
\end{align}

		\bibliographystyle{JHEP}
		\bibliography{index_D1-D5}

	\end{document}